# Informing research priorities for distributed acoustic sensing: insights from a citizens' panel

October 2026

SOUNDSCALE

University of Southampton
National Oceanography Centre
National Centre for Social Research
Ipsos

Funded by: UK Research and Innovation

***Informing research priorities for distributed acoustic sensing: insights from a citizens' panel***

Hannah Dalgleish[1,†,*], Rafael Mestre[1,†,*], Heidi Wheatley[1], Angela YT Chan[2], Mariya Yanishevskaya[3], Duncan Grimes[3], Sophie Hart[4,5], Sergey Astakhov[1], Matt Ryan[6], Dianna Smith[7], Alexandra Anikina[2], Ceri Davies-Tyrie[8], Gianluca Audone[1], Mohammad Belal[1,9], Louis Cook[1]

[1] *School of Electronics and Computer Science, University of Southampton*

[2] *Winchester School of Art, University of Southampton*

[3] *National Centre for Social Research (NatCen)*

[4] *Public Policy Southampton, University of Southampton*

[5] *Transportation Research Group, University of Southampton*

[6] *Department of Politics and International Relations, University of Southampton*

[7] *School of Geography and Environmental Science, University of Southampton*

[8] *Ipsos UK*

[9] *National Oceanography Centre, Southampton*

† *equal contribution*

* *corresponding addresses: h.dalgleish@soton.ac.uk; r.mestre@soton.ac.uk*

A work of the SOUNDSCALE project, funded by UK Research and Innovation (MR/Z505845/1).
https://soundscale.ac.uk

DOI: 10.5281/zenodo.23664714

# Table of Contents

# Executive summary

Remote sensing technologies, like space satellites or drones, acquire information about the Earth from afar. These technologies provide useful data for a wide range of applications, from modelling climate and weather, to monitoring crop and forest health, to managing disasters and humanitarian aid efforts, to optimising city operations and enhancing public safety.

As remote sensing technology advances and data are collected at ever higher resolution and over greater coverage, serious ethical and societal concerns emerge, relating to data privacy, surveillance, and the exacerbation of social inequities. These concerns are reflected in a 2022 report on public opinion on emerging technology:[1]

- People feel uninformed and unable to influence how their personal data is used. The systems set in place to govern such technologies are generally seen as untrustworthy.
- Technology needs to benefit and be shared equitably across populations and should not disadvantage any demographic.
- The use of emerging technology needs to be safe, secure, and responsible – people's identities, health, and wellbeing need to be protected.

As part of the research project SOUNDSCALE (Sensing On Urban Noise: Distributed Sensing for Collaborative and sustainable cityscapes And Living Environments), this report sheds light on an emerging remote sensing technology called distributed acoustic sensing (DAS). DAS turns ordinary fibre-optic cables into sensors which can detect and record environmental vibrations. As part of our research, we are using the National Dark Fibre Facility (NDFF) to collect DAS data using a fibre-optic cable between Southampton and London.

SOUNDSCALE is investigating the development of DAS technology using place-based and inter- and transdisciplinary approaches. To inform and develop our research for the SOUNDSCALE project, we wanted a method of engagement that would empower citizens, giving them a voice and a meaningful role in shaping the direction of our work. We convened a Citizens' Panel made up of 48 residents from Hampshire and London, living within 2.0 km and 0.5 km, respectively, of the NDFF fibre-optic cable. The Panel met during May and June 2025, combining online and in-person deliberation. Over three workshops, we gleaned the views, priorities, and concerns of residents about this emerging technology.

The SOUNDSCALE team is transdisciplinary, made up of academics and practitioners dedicated to the democratisation of science. By integrating the public into our research, we aim to prioritise and anticipate issues to mitigate possible ethical and societal repercussions of DAS. Our research tries to ensure that emerging technologies like DAS are deployed in a sustainable, just, inclusive, and equitable manner, with appropriate regulation. At the same time, we are examining the process of creating and producing citizen-informed interdisciplinary research. In this report we reflect on the role of the Citizens' Panel as a means for democratising academic research by increasing participation and citizens' control over how new technologies are researched, developed and deployed.

## Key insights from the Panel deliberations

After 11 hours of online and in-person workshops, the participants expressed intrigue, cautious interest, and uncertainty about DAS. The Panel often wanted research on DAS to focus on areas which they believed to be for the "greater good", such as improving health or protecting the environment. On the other hand, they tended to show apathy or anxiety toward prioritising research relating to threats, such as the use of DAS for military or surveillance applications.

[1] British Science Association, *Tomorrow's Tech, Today,* 15 June 2022, britishscienceassociation.org/news/public-opinion-on-emerging-technology-new-report

In total, five thematic areas of interest emerged through the deliberative process: environment; city infrastructure; health and community; surveillance and crime; and governance and costs. Although we report how participants later voted on priorities for immediate research, we place great emphasis here on the deliberative discussions themselves to illustrate public reasoning and concern:

- **Environment** was frequently prioritised as a further research area, particularly in Hampshire, because it was easy for citizens to agree that climate change and ecological protection are primary concerns and that any environmental application would benefit the "public good". Discussions about the environment did not focus only on environmental monitoring, but included questioning the potential environmental footprint of DAS, such as the environmental costs of data collection, storage, and analysis.
- **City infrastructure** was seen to be the most intuitive application of DAS, particularly for monitoring traffic, utilities, and maintenance. However, as understanding of the technology deepened, participants became more critical. They questioned whether using DAS for infrastructure applications was fully warranted and would add value to existing smart city technologies. While citizens considered it to be the most sensible or tangible application, city infrastructure was often deprioritised as it was either thought to be redundant or less beneficial compared to other research areas.
- **Health and community** prompted interest in participants, especially on a personal level. While the panellists were not explicitly told DAS could have health applications, they showed interest in investigating any direct applications for health. They acknowledged difficulty in seeing how vibration-based sensing could translate into health benefits, but were driven by a desire for new medical applications and uses to be developed. Some participants highlighted potential uses for public health, such as monitoring noise pollution.
- **Surveillance and crime** generated substantial discussions about pervasive monitoring, loss of privacy, hacking, state overreach or misuse of DAS. Discussions were accompanied by an apparent loss of agency on the topic of surveillance, questioning whether this technology adds anything new to existing surveillance systems, combined with the feeling that more surveillance is unnecessary. Despite these concerns, this theme was often deprioritised as an area of further research, suggesting fatigue or ambivalence towards the topic, or participants interpreting "priority" as support for developing surveillance applications, rather than research into safeguards, acceptable uses, or the prevention of harm.
- **Governance and costs** appeared across all discussions. Participants often asked about who controls and regulates the optical fibre infrastructure and the data itself, how decisions are made, the role of industry and powerful corporations, and whether public funds should be directed towards DAS given other pre-existing technologies, especially in the context of other urgent societal challenges.

At the very end of the Panel, participants were given one vote to select their favourite idea or research question which they had come up with and refined throughout the course of the final in-person workshop. Across the Panel as a whole, governance and costs received the highest number of final starred votes (13 London, 6 Hampshire), where participants prioritised monitoring and regulating DAS as it evolves, assessing its costs and benefits, and considering the consequences of not developing it. Environmental priorities were prominent in Hampshire (10 votes)—particularly for protecting the planet and mitigating the climate crisis—but not in London (0 votes). Voting did not capture the full reasoning behind participants' choices. Participants interpreted "priority" in different ways, and their views often changed as they learned more, heard from others, and considered trade-offs; the votes therefore need to be interpreted alongside the deliberations.

# Deliberating on an unfamiliar technology

A central question of the project was how to deliberate on an unfamiliar, emerging technology. Participants needed time to develop a basic understanding of DAS before they could engage confidently with the topic, and periods of partial understanding were sometimes characterised by discomfort, low enjoyment, scepticism, or suspicion.

To minimise overwhelming the participants with too much information, the research team slowly introduced the technology by first presenting a basic explanation of DAS with example use cases in maritime settings. This way we could provide recognisable and tangible examples, while trying to avoid influencing or priming the participants on potential applications of DAS in cities. While limiting information on land-based DAS applications may have hindered their understanding, instead we found that the participants quickly produced relevant research ideas, including topics which commercial or academic researchers are already investigating.

As expected for a relatively unknown emerging technology, a minority of participants struggled with key concepts about how DAS works or functions. They made proactive attempts to understand DAS by drawing upon concepts like microphones or the human ear. The level of comprehension adds another layer of complexity to the deliberative process, where some panellists may prioritise research areas differently according to their understanding of DAS. Although the technology does not measure sound, mechanical misunderstandings did not prevent participants from identifying relevant research areas and raising substantive questions about its possible uses and consequences. Later we reflect in more detail on learnings about engaging the public with lesser-known technologies at a lower technological readiness level than practitioners might be used to.

Uncertainty was not limited to the technology itself. Several participants initially found it difficult to understand their role in the Citizens' Panel, questioning their contribution as non-experts, especially when the task was to provide feedback on a technology that was still emerging. In a few cases, this translated into suspicion about hidden agendas, commercial interests embedded in the Panel, or undisclosed government involvement. We believe that this may be a symptom of both the structure of deliberative work and the technological topic. Scientific explanations alone are not enough to mitigate feelings of suspicion or mistrust and need to be accompanied by long-term community engagement. Therefore, taking the time to create spaces where citizens can explore suspicion and attitudes of (dis)trust in deliberations on technology is crucial, as contributing to debates once technologies are closer to application does not necessarily solve problems of trust. Artistic research, as part of the SOUNDSCALE project and taking place later in the two-year span, can be consulted as an active exploration of the themes of community engagement on these topics.

Clearer communications about the aims and purpose of the SOUNDSCALE project in the second workshop partly helped to mitigate suspicion and empower the panellists. We argue that where confusion or scepticism lingered, the inclusion of less trusting citizens, who are usually more difficult to recruit, is a good indication of the validity, diversity, and quality of the Panel. This is reinforced by very low participant attrition during the Panel, where only two out of fifty recruits did not attend the third workshop due to calendar conflicts. Our reflections suggest that future consultations on emerging technologies should explicitly emphasise the importance of early input, both to mitigate path dependencies and to make visible the role of academic and non-profit research. We also recommend drop-ins with researchers to understand issues and build trust.

# Reflections and learnings from the SOUNDSCALE team

The Citizens' Panel achieved a very high level of participation, with 48 of the 50 recruited panellists completing all three workshops, compared to a target of 40. Participants asked critical questions, exchanged views, revised their priorities over time, and identified research directions relevant to the SOUNDSCALE project. Despite efforts to ensure representation during recruitment, the achieved sample underrepresented some groups, particularly people from lower social grades and, in Hampshire, ethnic minority groups. Although a well-known challenge in deliberative activities, organisers should dedicate additional time to finding innovative ways to recruit from underrepresented groups, perhaps through wider community engagement.

One key lesson from the Panel is that deliberating on unfamiliar early-stage and complex technologies is not simply a matter of giving participants more information. Gradually introducing DAS through tangible maritime examples was useful, even when these came from a different application context, as it gave participants some solid footing to begin forming their own questions without becoming fixed on predetermined urban use cases too early. Participants developed a basic understanding of DAS through a brochure, a world café with researchers, repeated opportunities for questions, and ongoing dialogue with project investigators. This allowed participants to test their understanding, envisage possible applications, and refine their priorities over time. Creative methods such as scenario-building also helped participants engage with an unfamiliar technology by enabling them to explore possible futures and their implications without requiring technical expertise. However, splitting information across workshops left some participants with partial information in between sessions, which a minority interpreted as evidence of a hidden agenda. Phased information provision can support open-ended deliberation, but if there is too long a gap between stages, partial understandings or suspicions may develop before they can be addressed. Future panels on unfamiliar technologies, especially those at the research stage, should combine tangible examples and phased information with greater continuity between sessions, clear explanations of the research context and process, and sustained opportunities for dialogue with researchers.

At the same time, the Panel showed the importance of being clear about the role participants are asked to play: they are not expected to become technical or academic experts, but to bring their values, concerns, local knowledge, lived experience, and questions into the research process. Since DAS remains far from widespread application, it was difficult for participants to see how their input might be reflected in practice, and informing academic research was unfamiliar compared to commercial market research or consultation on an existing product. Future panels may therefore benefit from a short explanation of what a research agenda is, how academic research is funded, developed, and conducted, what sits inside or outside the scope of the process, and how the research team will translate participants' concerns into research priorities without narrowing those contributions into academic categories too early. Although these points were briefly explained at the beginning, more time was needed due to the unfamiliar nature of the Panel format and topic. Moreover, some participants understood "priority" as a desirable application, while others prioritised according to feasibility, personal importance, or public benefit. Overall, while voting offered a simple way to capture impressions at different points, the richest material came from the deliberations: the exchange of views, the testing of ideas, the role of researcher dialogue, and the ways participants' priorities changed as the process developed.

# How to read this report

The purpose of the Citizens' Panel was to determine key priorities for informing academic research on DAS. The quotations in this report have been drawn from the transcripts of the workshops and are used to amplify the voices of the panellists. Some quotes have been edited to improve readability, but there have been no additions nor word replacements which would change meaning or sentiment.

If you are interested in learning about emerging and remote sensing technologies:

- Information about DAS is described on page 9 followed by some of the implications of the technology summarised on page 10.
- The Panel's deliberations are summarised on pages 21 to 35. These are essential for understanding the concerns and priorities of citizens regarding remote sensing technologies.

If you are interested in the process and evaluation of the Citizens' Panel:

- The methodology describes the project design on pages 14 to 20.
- Our reflections on pages 40 to 47 summarise some of the challenges and successes for using deliberative methods to inform academic research.
- The Appendix describes in more detail our ethics approval; Panel recruitment and sampling; the planned and eventual quotas for the Panel; and the final research priorities put forward by the Panel for voting.

# What is DAS?

Distributed acoustic sensing (DAS) is an emerging remote sensing technology that turns ordinary fibre-optic cables into sensors which can detect environmental vibrations. Instead of installing an array of instruments along a route, DAS can use a single cable to record movement and vibration occurring nearby, such as traffic, construction activity, pipe leakages, or seismic motions.

A DAS interrogation unit sends optical pulses down a fibre-optic cable. The interrogator records the portion of light that naturally bounces back from minute imperfections in the fibre, a phenomenon called Rayleigh backscatter (Figure 1). Environmental vibrations cause minute stretches or compressions in the cable. These microscopic changes alter the phase pattern of the backscattered light (where the phase is a physical property of the light). By tracking how the phase of the backscattered signal changes over time, the system can track disturbances along the cable.

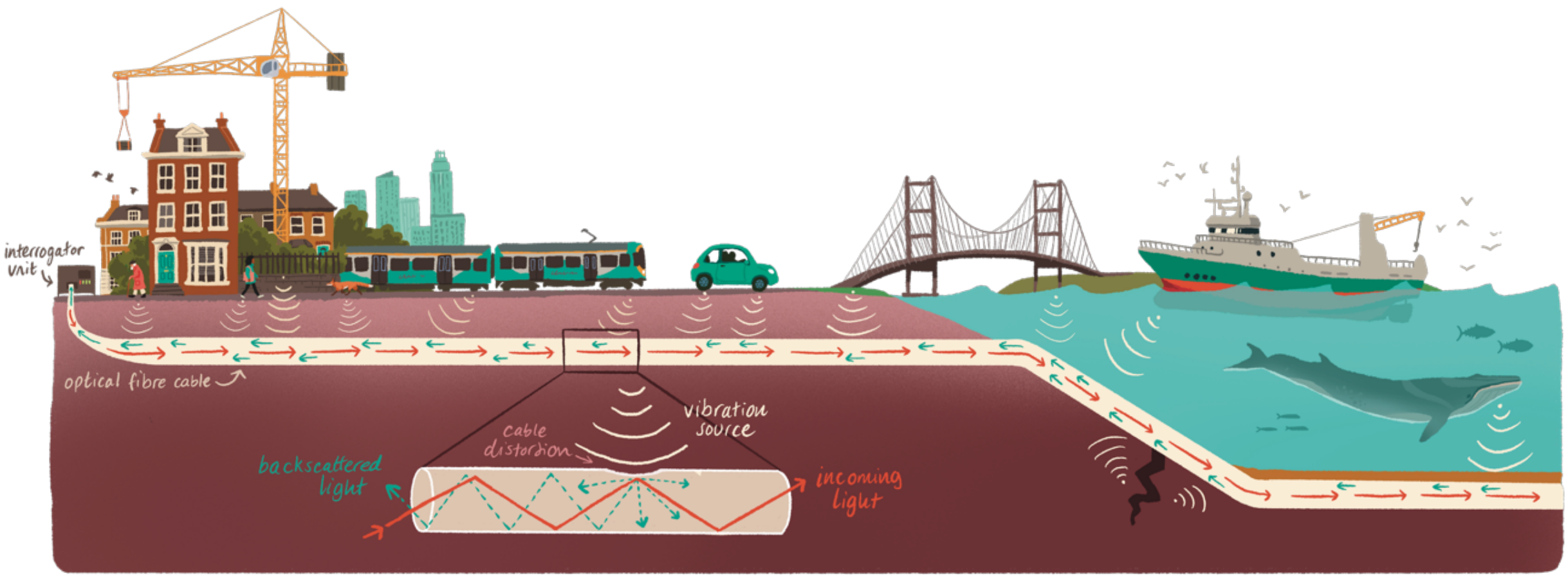

**Figure 1:** Schematic of distributed acoustic sensing (DAS) on land and at sea. Environmental vibrations create minute distortions of the fibre-optic cable, leaving an imprint on the backscattered light. These effects can be detected remotely and in real-time. Credit: Dominique Vassie.

In practice, DAS can effectively repurpose unused fibre telecommunication infrastructure underground, known as "dark fibres". This creates a continuous line of virtual sensors along the cable length. The data collected can not only be used to detect events, but also to estimate their location and, in many cases, distinguish between different activities based on their characteristic vibration patterns, for example, a moving car versus a falling tree. The application of remote sensing with fibre-optic cables offers several advantages compared to traditional sensors, as it is robust against electromagnetic interference, adverse weather, and low visibility conditions.

DAS has so far been used in various applications, such as monitoring pipelines, tracking whales and ships, or detecting earthquakes and avalanches.[2] A recent BBC article describes how DAS was used to discover over 100 water leaks in England, saving two million litres of water per day.[3] DAS also has military and security applications, including in commercial contexts.[4] As the technology continues to advance, DAS could potentially address a range of urban challenges or environmental health issues, perhaps by improving smart city infrastructure for better transport planning or monitoring and mitigating noise pollution, for example.

[2] Landrø, M., *et al.*, *Sensing whales, storms, ships and earthquakes using an Arctic fibre optic cable*, Scientific Reports, 12(1), 19226 (2022). DOI: 10.1038/s41598-022-23606-x.
[3] Stallard, E., & Tanaka, M., *The race to stop England running out of water*, BBC News, 29 August 2026, bbc.co.uk/news/articles/clyrv84yd9ko.
[4] Gato Security, *Top applications of distributed acoustic sensing in security*, 12 June 2026, gato-security.com/top-applications-of-distributed-acoustic-sensing-in-security/.

# Rationale: possible DAS risks and consequences

Governments face increased pressure to respond to complex urban challenges such as climate change, decaying infrastructure, and public safety. All these challenges require extensive and accurate datasets for monitoring and modelling solutions. DAS could potentially offer large-scale and cost-effective sensing capabilities that do not require intensive deployment or maintenance compared to other remote sensing technologies like surveillance cameras, environmental sensors, or arrays of connected Internet of Things (IoT) technologies.

The repurposing of existing and invisible underground telecommunication infrastructure to sense the environment raises certain social and ethical questions, however. Who can collect and use DAS data? Could personal information be obtained without the person knowing? Could DAS data perpetuate bias and result in unfair assumptions about individuals or communities? How is DAS data stored and shared and how costly is it, environmentally and economically? Responsible research and deployment of DAS require clear governance, transparency about capabilities and limitations, and careful consideration of potential consequences. DAS carries a risk of compromising individual privacy over time in addition to environmental costs. One day of raw DAS data from a 50 km fibre can require around 85 terabytes of storage, although temporal or spatial downsampling can reduce this. Processing and analysis also produce carbon emissions.

In the absence of safeguards, DAS could exacerbate existing social inequalities and erode public perceptions of trust. Amnesty International found that the "inherent bias" in structural and institutional racism and discrimination is reflected in the historical policing datasets used in predictive policing tools, such as geographic, individual, profiling and risk assessment predictive policing systems. The police forces' systemically disproportionate targeting of Black and racialised people, and people from lower socio-economic backgrounds, not only "leads to their increased criminalisation, punishment, and exposure to violent policing," but is also exacerbated by algorithmic models and non-transparent uses of these technologies.[5]

Limited public input and an absence of dedicated regulatory frameworks are symptomatic of tendencies for emergent technologies to be utilised ahead of comprehensive governance and citizen input. The lack of public consultation and regulation surrounding facial recognition technologies, for example, has been subject to sustained legal challenge, civil society campaigning, and public concern about privacy, misidentification, racial bias, and the expansion of police and commercial surveillance.[6] These concerns have prompted reactive development of regulatory frameworks and guidance to address discriminatory shortfalls.[7] For example, South Wales Police's use of automated facial recognition was ruled unlawful by the Court of Appeal in 2020,[8] and Essex Police paused the use of live facial recognition after a study identified potential accuracy and racial bias risks.[9] We hope to contribute to the research addressing societal and ethical issues that DAS may create at an early stage of its development.

---

[5] Amnesty International UK, *Automated Racism: How police data and algorithms code discrimination into policing*, 1 February 2025, amnesty.org.uk/knowledge-hub/all-resources/automated-racism-report/.
[6] Big Brother Watch, *Stop Facial Recognition*, accessed 6 July 2026, bigbrotherwatch.org.uk/campaigns/stop-facial-recognition/.
[7] Ada Lovelace Institute, *Beyond face value: public attitudes to facial recognition technology*, 2 September 2019, adalovelaceinstitute.org/report/beyond-face-value-public-attitudes-to-facial-recognition-technology/.
[8] Jenny Rees, *Facial recognition use by South Wales Police ruled unlawful*, BBC Wales, 11 August 2020. bbc.com/news/uk-wales-53734716.
[9] Robert Booth and Mark Wilding, *Essex police pause facial recognition camera use after study finds racial bias*, The Guardian, 19 March 2026, theguardian.com/technology/2026/mar/19/essex-police-pause-facial-recognition-camera-use-study-racial-bias.

# About SOUNDSCALE

We are SOUNDSCALE, an inter- and transdisciplinary research team. We follow a citizen-centric, justice-centric and bottom-up approach, dedicated to the responsible research and development of an emerging technology called distributed acoustic sensing (DAS).

Our team includes researchers from the University of Southampton, National Oceanography Centre (NOC), National Centre for Social Research (NatCen) and Ipsos UK. SOUNDSCALE is funded between Jan 2025–Mar 2027 by UK Research and Innovation (UKRI), the national funding agency investing in science and research in the UK. Our project is divided into three overlapping phases:

1. **Data collection from the Citizens' Panel and DAS**. The DAS data are acquired using a fibre-optic cable belonging to the National Dark Fibre Facility (NDFF), a UKRI facility, which connects several cities in the UK, including Southampton and London.
2. **The development of transdisciplinary research**, informed by the Citizens' Panel and an interdisciplinary workshop in September 2025. The workshop helped to translate the Citizens' Panel priorities into actionable research areas, fostering innovative methodologies and novel interdisciplinary knowledge. Research on DAS, spanning the physical and mathematical sciences, social sciences and artistic investigations, is ongoing.
3. **Synthesising findings for engagement with policymakers and the public**, ensuring that the research benefits are tangible and aligned with societal needs. Further engagement with the public includes community workshops and exhibitions about public infrastructure. These artistic and community-facing activities are intended to extend the questions raised by the Citizens' Panel, particularly around public infrastructure, place, and how people understand technologies that are embedded in everyday environments.

The challenges, opportunities, and questions new technology poses can only be adequately addressed by combining the domains and understandings in which the technology itself is created and embedded. Thus, SOUNDSCALE delves into and integrates social, physical, political, geographical, technological, environmental, and arts-based research. Through knowledge exchange and policy and community engagement, we co-create between academics and non-academics, including practitioners, activists, artists, policymakers, community groups, and individual citizens. This collaborative philosophy is designed to produce research that is not only academically rigorous but also more attentive to social concerns, public priorities, and the possible consequences of DAS as it develops.

We believe that DAS must benefit society and that the potential harms and risks of this emerging technology need to be appropriately considered. Research should be more democratic, and so we have taken a citizen- and justice-informed approach to ensure that our research aligns with public expectations and societal priorities. Given the relative infancy of DAS, citizen attitudes, expectations, and accepted use cases remain underexplored. To address this, we convened the Citizens' Panel to explore perspectives surrounding DAS and to identify priority research areas.

# Why a Citizens' Panel?

Citizens ought to have a say on emerging technologies which could significantly impact society. It was therefore important for the SOUNDSCALE project to empower citizens, giving them a voice and a meaningful role in shaping the direction of our research. The participants would need time and space to learn about the unfamiliar and complex topic of DAS in a guided environment to develop their views and identify relevant priorities and questions.

Deliberative methods were well suited to these aims. They offer participants the time, information and discursive conditions needed to enable respectful, reflective, and well-reasoned in-depth discussion. These methods have already been successfully used to explore scientific and technical policy questions, often through Public Dialogue such as the Sciencewise programme. They are important in enabling informed discussion and uncovering not just what matters to the public about science and technology issues, but why it matters. Making science more democratic can also increase scientific literacy, strengthen trust in publicly funded research, and reinforce the role of science as a tool for the common good.[10]

We chose a Citizens' Panel format because the SOUNDSCALE project needed more than immediate reactions to an emerging technology. The Panel enabled participants to learn about DAS, discuss uncertainties and trade-offs, hear from others, and express and revisit their views before identifying priorities for research. We were also aware of the limitations of Citizens' Panels, particularly when used to deliberate on unfamiliar and early-stage technologies. Participants may need time to understand both the technology and the purpose of the process, and a panel cannot by itself resolve wider issues of trust, uncertainty, or unequal expertise. We tried to address these limitations through a staged design: combining accessible explanations of DAS, repeated opportunities for questions, small-group deliberation, interaction with researchers, and chances for participants to revisit their priorities as their understanding developed. We return to these design choices and their strengths and limitations in Reflections on the Panel process.

Our choice of a Citizens' Panel also shaped the kind of deliberation that was possible. The Panel was explicitly not designed around a settled policy choice, or around presenting participants with clearly defined "for" and "against" positions on DAS. Given the early stage of the technology and the purpose of SOUNDSCALE, many of the possible applications, risks, and trade-offs were still uncertain. The aim was therefore not to produce a rounded measure of public opinion on DAS, but to support panellists to explore an unfamiliar technology, raise questions, and help inform research priorities within the scope of the project.

We convened the Panel in two geographic areas where the dark-fibre cables relevant to our research are located through the NDFF: London and South Hampshire (Southampton-Eastleigh-Winchester). This place-based approach served three purposes. First, it grounded the work in people's lives and locations, enabling participants to reflect on how DAS might affect the communities and spaces they know best. Second, it aligned the engagement process with the locations where SOUNDSCALE is collecting DAS data, ensuring that the perspectives informing the research came from people living close to the infrastructure being studied. Third, it enabled us to create two local panels that could be re-engaged over time, drawing upon participants' knowledge and lived experience to determine priorities in a considered way. The panellists can then see how their recommendations are incorporated as the research develops, for instance through further community workshops through our artistic research strand. We plan to re-engage with panellists who consented to being re-contacted to discuss how the project's research has progressed since the Panel and to seek their input. Bringing together participants from two different locations also allows for the mixing of diverse viewpoints and experiences, creating opportunities for new, or combinations of, ideas to emerge.

[10] Mocanu, M., Bibiri, AD., Rusu, V.D. *et al., Enhancing civic engagement with science: a comparative approach across European regions*, Scientometrics, 130, 447–468 (2025). DOI: 10.1007/s11192-024-05198-7.

The Citizens' Panel was also an opportunity for the SOUNDSCALE team to explore alternative ways to democratise expertise. Deliberative methods include input from experts to ensure that participants have access to factual information to minimise the risk of misunderstandings or uninformed biases dominating the outcomes. A diverse group of experts will usually be selected to demonstrate competing perspectives, so that no one view dominates. The inclusion of experts can introduce power imbalances, however, where top-down narratives are validated and quiet voices can become even quieter. We drew on the diverse expertise within the SOUNDSCALE team, who tried to remain as neutral as possible. While most of us were not technical experts in the technology itself, we offered more critical insights about how the technology was relevant to our areas of expertise or posed questions for the panellists to consider. We also invited two 'lay experts', people who had previous experience deliberating as participants in a panel on a different topic, and we asked them to introduce DAS to the participants. We provided these participants with some structured support to inform themselves about DAS. This was done to understand how learning foregrounded by peer experts compared with learning from career researchers.

The Panel followed a hybrid approach and comprised three sessions in total. The first two sessions were online and lasted 2.5 hours each, with participants from both locations together. This enabled participants from London and Hampshire to learn about DAS at the same time, hear from the research team together, and begin developing a shared understanding of the project. The third and final session was a full-day in-person event, held simultaneously in Southampton and London, lasting 6 hours. We chose this format because the online workshops were well suited to introductory learning and cross-location engagement, while the in-person workshop allowed more time for complex deliberation, place-based activities, and finalising research priorities once participants were more familiar with the topic.

The recruitment and design of the workshops for the Citizens' Panel were reviewed and approved by NatCen's internal research ethics committee (REC). Additional ethics approval was obtained by the SOUNDSCALE team from the University of Southampton's Faculty of Engineering and Physical Sciences REC (ERGO/FEPS/104481.A1). See Research ethics for more details.

# Panel design and delivery

The Citizens' Panel consisted of 48 members of the public from Hampshire and London who participated in three workshops delivered between May and June 2025. Trained facilitators guided the participants as they deliberated on information about DAS to develop their views, concerns, and priorities about this emerging technology.

Quotas were set to recruit a "mini-public"[11] (a sample matching the important demographics and attitudinal diversity of a broader population), reflective of the two areas of interest (Hampshire and London). As a place-based Citizens' Panel, we chose as geographical boundaries for recruitment the distance from residence to the [National Dark Fibre Facility]() optical fibre infrastructure, as participants might be able to relate more closely to the research process. We selected postcodes within 2 km of the fibre in Hampshire (Southampton-Eastleigh-Winchester) and 0.5 km in London (East and West), where population density is much higher. Central London boroughs were excluded due to high concentrations of non-residential areas.

From these postcodes, we recruited a diverse group of panel members across age, gender, social grade, ethnicity, education, disability, and long-term illness. Further quotas were increased for those from lower social grades (through indices of multiple deprivation metrics), as their views are often underrepresented in social research. We also included quotas for a range of worldviews using a spread of political leanings towards major parties, in line with best practice. As far as possible, we tried to reflect the demographics of those living near the cable. To ensure we could more clearly attribute any differences between London and Hampshire discussions to geography rather than selection bias, we started by reflecting demographics in Southampton and then selected participants in London with matching characteristics to those selected in Hampshire. Anyone with prior knowledge of fibre optics and/or distributed fibre-optic sensing technologies was excluded from taking part.

We set a target sample size of 40 participants, and a total of 48 participants completed all three workshops, 20% greater than the target sample size. Overall, the achieved sample reflected the Southampton population in key demographic characteristics like gender, age, disability, and included a spread of different political perspectives. There were marginally more women who attended than men, and there was an overrepresentation of 45–64-year-olds across both Hampshire and London. In Hampshire, desired quotas were not achieved for any ethnic minority groups, creating an overrepresentation of participants who identified as white. In both locations, quotas were not met for participants with no qualifications, and participants within AB and C1 social grades were overrepresented compared to C2 and DE.[12] While the London sample achieved quotas for the Indices of Multiple Deprivation (IMD),[13] the Hampshire sample did not meet quotas for deciles 1 and 2 (highest deprivation) as the locations bounded by closeness to the fibre included few or no areas in those deciles. See Panel recruitment, Panel sampling, and Quotas in the Appendix for more details.

Fifty panellists were paid £220 each for participating in the Panel. Other incentive-related costs and expenses totalled £736.80. Recruitment and project management costs, including VAT on incentives, came to £7,960. Catering (in London and Southampton) and venue hire (London only) costs for the in-person workshop were £3,843; the cost of the Southampton venue was in-kind by the University of Southampton. The lay experts were remunerated £250 in total for their preparation for and presentation at the first workshop. The direct costs of the Citizens' Panel totalled £23,789.80 (including VAT). This excludes staff costs for facilitation and other support. Although some staff costs are directly traceable because they were paid by assignment, much of the work was provided through in-kind or directly allocated time from the SOUNDSCALE team, making the total difficult to estimate. We therefore report only direct non-staff costs.

---

[11] Escobar, O., & Elstub, S., *Forms of Mini-publics: an introduction to deliberative innovations in democratic practice*, New Democracy, 2017, newdemocracy.com.au/forms-of-mini-publics.
[12] See ons.gov.uk/census/aboutcensus/censusproducts/approximatedsocialgradedata for a definition of social grades.
[13] See ocsi.uk/2025/10/10/what-are-the-english-indices-of-deprivation-a-beginners-guide-to-the-imd/ for a definition of IMD.

The first two workshops were 2.5 hours each and held online on weekday evenings, two weeks apart. In these workshops the participants learned about DAS, explored its potential uses, and generated initial areas of interest. The final in-person workshop took place on a Saturday in London and Southampton simultaneously, where participants were given a full day (6 hours) to consider and develop their priorities further.

All workshops involved a mix of plenary and breakout group sessions. As well as learning about DAS, the panellists learned about the research team and our different disciplinary backgrounds, as well as the purpose and motivation of our research project. Discussions and activities were moderated by facilitators in smaller groups of 7-9 people. The small groups were separated according to location (London versus Hampshire). All breakout discussions were recorded to enable data analysis after the Panel, and the facilitators took notes and met after each workshop to share reflections about their group, note areas of concern, and provide feedback for improvement. Table 1 summarises the Panel and its main activities.

**Table 1:** Citizens' Panel activities across the three workshops.

| Workshop | Format | Main activities |
|---|---|---|
| **1** | Online | 1. Local issues and views on tech<br>2. Intro to DAS<br>3. DAS applications to local issues<br>4. Prioritising areas of interest |
| **2** | Online | 1. Intro to team and research project<br>2. Coming up with priorities and questions<br>3. World café with research team on five themes<br>4. Reassess priorities |
| **3** | In-person (London and Southampton simultaneously) | 1. Review priorities<br>2. Sci-fi world-building exercise<br>3. Risks and opportunities for DAS<br>4. Place-making: choice between walk or map<br>5. Finalising priorities (voting) |

A short video documenting the Citizens' Panel and its activities is available on YouTube: https://youtu.be/lKTVPRI7JM0.

# Workshop 1

The aim of the first workshop was to introduce participants to our project and DAS technology. After a quick round of introductions from NatCen and the SOUNDSCALE team, the participants were moved into breakout groups. To set the scene, participants shared how optimistic or pessimistic they felt about the future and whether technology would play a positive or negative role. They were then asked to identify issues in their local area that they felt passionate about, and how technology might be able to improve them. The discussions generated an initial list of topics and areas of interest for each breakout group that were documented by the facilitator in Miro (an interactive digital whiteboard). The participants also shared their hopes, concerns, and expectations about the Panel, before learning about DAS.

A secondary aim of the workshop was to explore how best to present technical information to participants, specifically any differences between information presented by 'experts' compared to members of the public. To explore this, participants were divided into two "mini-plenaries"—one group heard about DAS from the research team (DAS experts), and the other group heard from two citizens who had previously engaged in a public dialogue on research and development,[14] but were not part of this Citizens' Panel. After being briefed by a member of the research team, the lay experts were asked to research DAS in marine contexts to prepare a short 5-minute presentation about the topic and answer any follow-up questions. The two lay experts had never heard of DAS beforehand and they were reimbursed for their time. We introduced the Panel to DAS in marine settings to avoid priming the panellists with urban use cases.

After the DAS presentations, the panellists went back into their breakout rooms and explored how DAS might apply to some of the topics they had discussed earlier. Facilitators guided the participants to identify and prioritise their ideas which were shared with the whole Panel at the end.

In preparation for the second workshop, NatCen researchers analysed the data from the Miro board and identified six overarching research areas that participants were interested in: the environment, city infrastructure, community and equality, surveillance and crime, governance and costs, and health (see Figure 2).

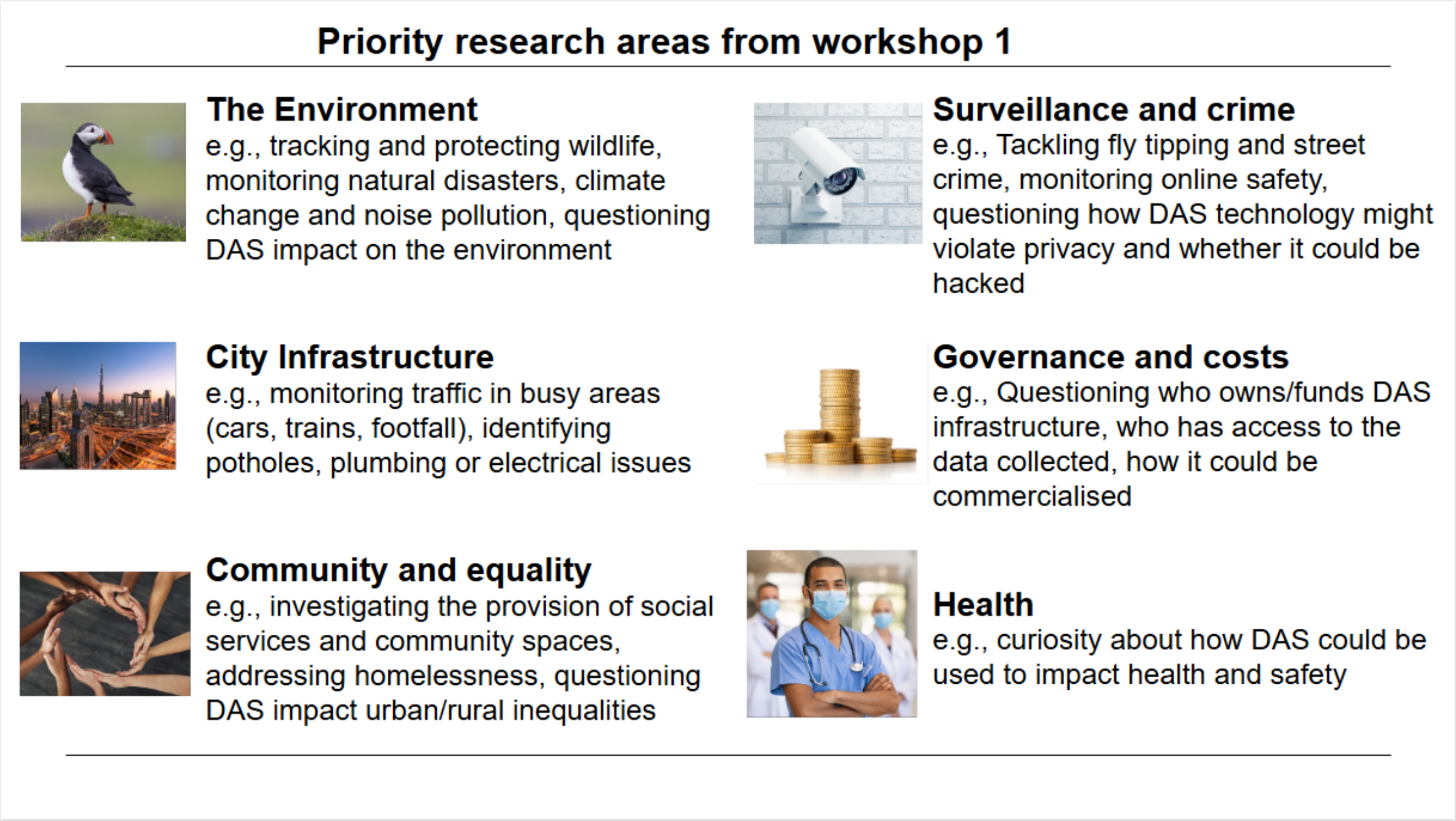

**Figure 2:** Emerging priority research areas from Workshop 1. 'Community and equality' and 'health' were combined for the following workshops.

[14] NatCen, *People's Vision for R&D: A public dialogue*, 9 September 2024, sciencecampaign.org.uk/analysis-and-publications/detail/peoples-vision-for-rd-natcen-and-nccpe-reports/.

# Workshop 2

Before the second workshop, the research team created and sent the panellists a brochure. Its purpose was to aid participants' understanding of DAS and their role by reiterating the motivation of the Citizens' Panel and outlining potential use cases of DAS in cities.

The second workshop was also online and provided participants with further information about the research project and some examples of research questions. In small groups, the participants discussed their understanding of the project aims and could ask a researcher questions about the scope and purpose of the project. Panellists were then invited to look at the research areas identified in Workshop 1 and asked to choose which area they would prioritise and why.

The second half of the workshop took on the format of a "world café". Five members of the research team, each an expert in one of the areas, rotated between the breakout groups to explore those priority areas in more depth. We combined 'health' and 'community and equality' to create one theme ('health and community'), due to the overlap between the two. After further questions and discussions, the participants were asked to choose their priority area again and share their reasoning as to why they had changed (or not changed) their view.

# Workshop 3

In the final workshop, participants were brought together in person to take part in a wide range of activities. The participants located in the South Hampshire area convened in Southampton, and those living in London in a central London venue. The final workshop was designed to move from understanding DAS as a technology towards reflecting on its possible futures and local implications. The panellists came up with further questions and ideas about the research areas they were interested in so that they could choose a set of final priorities by the end of the workshop.

The morning revolved around a science-fiction writing-inspired scenario-building exercise in which participants imagined future worlds related to DAS.[15] Participants formed into groups according to interest in the five themes presented in Workshop 2. The activity involved creating a 2x2 matrix where one axis represented a possible future capability of DAS, while the other depicted a social or environmental situation. The participants created their matrix by choosing two items (one for each axis) from a large selection prepared by the researchers (see Table 2). Each axis, then, represented a choice between two extreme versions of the themes that would define the future scenario (e.g. an axis themed "noise pollution" could range from "extremely high" to "very low").

In small groups, they imagined a future scenario for each quadrant as defined by the intersection of the two axes. They were asked to discard the most "dystopian" and "utopian" scenarios and explore the remaining two worlds to allow for more nuanced debate. The research team rotated around the breakout groups to ask the panellists questions to further develop their scenarios and encourage new ways of understanding DAS. This scenario-building exercise was designed to help participants think beyond current applications of DAS and consider its longer-term social, environmental, and ethical implications.

In the afternoon, participants chose between two place-based activities to shape their priorities in relation to their locality: an outdoor walk close to the DAS cable or an indoor activity reflecting on a map of the NDFF cable in London or Hampshire. The place-based activities were intended to move discussion beyond abstract considerations of the technology and ground deliberation in the participants' lived experiences. Whether through map making or walking along the cable, panellists were encouraged to draw on local knowledge, reflect on how DAS might interact with places and communities, and consider the social and environmental implications of the technology.

[15] See foresight.unglobalpulse.net/blog/tools/2x2-matrix-scenario-building/ for a description of scenario building.

At several points throughout the day, participants generated ideas and questions on post-it notes related to the five overarching areas of interest and, using sticky dots, voted on the themes and questions they felt were most important to prioritise. The participants also wrote down their wider reflections and rationale for choosing their research priorities in workbooks.

At the end of the Citizens' Panel, participants were invited to share any additional ideas which emerged from the scenario-building and place-based exercises. To conclude, they voted for their final priority idea or question that they felt was most important for the SOUNDSCALE team to later explore.

**Table 2:** Imagined scenarios built using axes of DAS functionality versus an extreme sociotechnical future during Workshop 3.

| DAS capability | Sociotechnical future | Research area | Location |
|---|---|---|---|
| *How much DAS data is stored and on what time scale?*<br>full data storage forever ↔ no data stored | *Accountability of government surveillance systems*<br>transparent and audited ↔ no oversight or accountability | Surveillance and crime | Hampshire |
| *When is DAS operational?*<br>constantly ↔ rarely | *Eavesdropping and invasion of privacy*<br>extreme ↔ non-existent | Surveillance and crime | Hampshire |
| *When is DAS operational?*<br>constantly ↔ rarely | *Privacy and data protection*<br>highly protected ↔ non-existent | Surveillance and crime | Hampshire |
| *When is DAS operational?*<br>constantly ↔ rarely | *Privacy and data protection*<br>highly protected ↔ non-existent | Surveillance and crime | London |
| *When is DAS operational?*<br>constantly ↔ rarely | *Personal health*<br>high ↔ poor | Health and community | London |
| *When is DAS operational?*<br>constantly ↔ rarely | *Noise pollution*<br>extreme ↔ non-existent | Health and community | Hampshire |
| *How well can DAS detect and distinguish between sources?*<br>clearly resolved ↔ unclear | *Noise pollution*<br>extreme ↔ non-existent | Health and community | London |
| *Can DAS data be accessed?*<br>open access ↔ no access | *Health knowledge and data*<br>comprehensive ↔ limited | Health and community | London |
| *How is DAS automated?*<br>fully ↔ humans decide | *Life expectancy*<br>high ↔ low | Health and community | Hampshire |
| *How sensitive is DAS?*<br>high ↔ low | *Sleep*<br>plentiful ↔ restricted | Health and community | Hampshire |
| *How well can DAS detect and distinguish between sources?*<br>clearly resolved ↔ unclear | *Agriculture and food production*<br>extreme ↔ non-existent | Environment | Hampshire |
| *How well can DAS can detect and distinguish between sources?* | *Climate change*<br>extreme ↔ reversed | Environment | Hampshire |
| clearly resolved ↔ unclear | | | |
| *How well DAS data is interpreted and used?*<br>beneficial ↔ misused or overlooked | *Animal habitats*<br>species unknown ↔ species accounted for | Environment | Hampshire |
| *How sensitive is DAS?*<br>high ↔ low | *Ecosystem quality*<br>healthy biodiverse ↔ degraded, high extinction of species | Environment | Hampshire |
| *How sensitive is DAS?*<br>high ↔ low | *Data accessibility*<br>open access ↔ no access | Governance and costs | London |
| *How is DAS integrated?*<br>fully connected ↔ operates alone | *Data accessibility*<br>open access ↔ no access | Governance and costs | London |
| *How widespread is DAS?*<br>high coverage ↔ low coverage | *Legislation*<br>overregulated ↔ no regulation | Governance and costs | Hampshire |
| *How smart is DAS?*<br>intelligent ↔ passive/dumb | *Legislation*<br>totalitarian ↔ laissez-faire | Governance and costs | London |
| *How widespread is DAS?*<br>high coverage ↔ low coverage | *Criminal activity*<br>all crime investigated ↔ no investigation | Infrastructure | London |

# The Panel deliberations

Across the three workshops, the panellists learned about DAS: how the technology works, how it could be used in different contexts, and some possible implications according to the five themes which emerged. Based on this information, the Panel members deliberated on their priorities for what research ought to be carried out on DAS. In presenting these deliberations, we focus not only on what participants prioritised, but also on how they reasoned through different possibilities, exchanged views, and sometimes changed their opinions as the Panel progressed. We do not see the value of the Panel as pertaining only to the final votes, but also in the way participants tested ideas, compared DAS with existing technologies, weighed public benefit against risks and costs, and reflected on the views of others.

We analysed the transcripts of the Panel deliberations, taking a top-down approach to coding based on the five themes which emerged in Workshop 2 (see Figure 2), in addition to a sixth theme on 'initial reactions to DAS'. The transcripts were coded and interpreted by five different researchers from the SOUNDSCALE team. We summarise the deliberations in the following subsections to share an in-depth representation of the participants' concerns and perspectives regarding DAS.

## Views on technology and local issues

At the beginning of the first workshop, we asked the panellists about their attitudes towards technology. Overall, people were optimistic and saw technology as having a positive impact on society and their lives (or having the potential to). As one person remarked, *"Technology for me has always been really positive [...] it does have its ups and downs, but [...] the things we can do with it is amazing and I can only see it getting better."* Others were more cautious, and a few participants saw technology as a double-edged sword, that both benefits and harms: *"[Technology is] not a panacea. It can't solve homelessness, can it? Or people taking drugs. How can it stop that? It's more about informing and educating people. But that's not the key driver, is it? It's to sell products."* Social media was also a useful example:

> *"It's all good at the beginning and then eventually it's not that great [...] it can be great but is it always going to be great? [...] Like social media [...] yes, it's great because you can keep in contact but in other ways it's bad."*

Some participants were worried about society becoming too dependent on technology, sometimes to the detriment of our health:

> *"Although I love technology and there is a place for it [...] I know how much my life relies on technology, but I'm also weary of to what extent do we stop being human and become robots."*

A smaller number expressed a sense of powerlessness about technological change, arguing that "*you can't stop it anyway. Really. I think you just have to accept the inevitable advance [of technology]*." Others were more sceptical, feeling helpless or very negative, believing that technology exists only to make profit, replacing people and taking away jobs (especially regarding AI): *"I think the issue we've all got here is that people use technology to make money. That's just the facts of life."* These concerns were frequently intertwined with distrust of government and industry:

> *"Electric vehicles are the worst idea that we've come up with, it's close to the nuclear bomb for [the] environment [...] if a lithium battery catches fire it keeps burning until the chemical reaction stops."*

> *"The government don't care about the environment at the end of the day […] HS2 took out so many acres of forestry […] when they're saying they care about the environment, I say load of tosh. Everything's to do with money at the end of the day."*

When we asked the Panel about their local interests and concerns, they highlighted a wide range of issues; the majority related to perceived problems or challenges in their local areas. Several themes cut across the issues identified by participants. First, concerns were often framed in terms of decline or loss, including the deterioration of public spaces, the loss of community cohesion, the closure of local services, and perceived failures of local governance. Second, many issues reflected tensions around technology itself. Participants raised concerns about AI replacing jobs or public services and about technology exacerbating existing social problems. Finally, many of the issues highlighted were interconnected, linking environmental quality, public health, infrastructure, governance, and community wellbeing. This interconnected understanding of local challenges was later reflected in participants' discussions of DAS, where the boundaries between themes frequently blurred. Table 3 summarises their local interests and concerns:

**Table 3:** Panellists' local interests and concerns.

| Theme | Positive context | Negative context |
|---|---|---|
| **Environment** | Wildlife; foxes; green spaces. | Climate change; noise pollution. |
| **Health and community** | Housing; local businesses. | AI taking over public services; homelessness; deterioration of local park; losing community closeness and cohesion. |
| **Surveillance and crime** | | Crime and safety; (cyber)bullying; antisocial behaviour. |
| **City infrastructure** | | Potholes; lack of cycle lanes and public transport; e-bikes as a public nuisance; overcrowding; poor connectivity and access to services; traffic; road works; redevelopments; fly tipping. |
| **Governance and costs** | | Mismanagement of council funds; cost of living; inflation; AI taking jobs; disability justice; closure of local businesses and public services. |

# Initial reactions to DAS

In the first workshop, the participants were asked about their hopes and expectations for the Panel and their initial understanding of DAS after a 5-minute presentation about the technology. Many immediately wanted to know what the realistic benefits of the technology would be, on personal and societal levels, and at what cost. As one participant put it: "*What's the value in [DAS] for me, you, society in general and at what cost? And how do you know it's proven?*" Another questioned, "*Is it actually going to be of any use and not just a big expense?*"

A few people were sceptical and pessimistic about DAS, unable to see how it could be used for good, questioning whether it addressed any genuine need and whether it duplicated pre-existing technologies. One participant commented that their "*initial feeling is [DAS solves] a problem that didn't exist.*" This was a feeling that was generally echoed by others: *"If I'm honest, I'm finding it quite hard to see how [DAS is] going to benefit people […] it doesn't seem like it's anything we haven't got."*

Even among those that were doubtful, there was often an openness to being persuaded if clear benefits could be demonstrated:

> *"I still don't understand this technology, how it's going to do things [...] I'm very mystified, very pessimistic at the moment. But I'm happy to be delighted with some amazing technology that can really, really help. But at the moment, I'm very pessimistic."*

Some panellists were curious and more positive about its potential use, but this was often framed around applications for safety:

> *"If there was technology that would make my home feel safer, [...] it would be a real benefit for me as an end user."*

> *"I'm interested to see if there will be a way to implement it for safety. I have no idea how they would do it, but that is interesting to me if they could."*

Alongside concerns about usefulness, many participants were cautious about unforeseen negative consequences of DAS and whether these were being considered or acted upon.

> *"It would be great to have a sense that people who are experts in this area have considered the positives of implementing a new technology, but also some of the risks around it, and some of the unintended consequences about the roll out of new technology. [I] hope that some of that had been considered and listened to from the outset to try and mitigate any of those potential risks."*

> *"My impression is that new technology will focus on the positives. And then 50 years later you realise, oh no, there have been all these negatives that we've not really identified."*

> *"I think sometimes when technology's introduced, the implications haven't fully been thought about, you know, ethical considerations, things like that [...] I would hope that with any new technology, particularly this one, that it'd be balanced, and all kinds of pros and cons have been shared and widely available for everyone."*

All groups noted the implications of DAS being used either for surveillance, war, espionage, or terrorism, in some way.

> *"I can't see [DAS] being used for the greater good. I think it could be used for the wrong purposes by hostile governments, quite frankly."*

> *"This sounds like, with all the best intentions, we are slowly sleepwalking into more surveillance."*

> *"Even though it's [ethical] to us as normal citizens, I think the use of [DAS for surveillance] would be classed as unethical. That would be a slippery road [...] it can encroach on your privacy."*

Over the course of the Citizens' Panel there was also a lot of uncertainty around the technology, with many questions asked about its role and how it functions. Participants often framed their responses in the form of questions, seeking to understand not only how the technology works, but also who controls it, who benefits from it, and what consequences it might have. The questions in Table 4, identified through analysis of the first workshop discussions, illustrate the breadth of issues that participants immediately associated with DAS and the types of information they felt they needed in order to form considered views about the technology:

**Table 4:** What the panellists wanted to know about DAS, summarised from Workshop 1.

| | |
|---|---|
| **Technical** | • How much sound does it pick up?<br>• Can it hear conversations?<br>• What can be detected and identified?<br>• Can sounds be differentiated?<br>• What is the range of DAS? How far can it detect from the cable?<br>• What does the data show?<br>• Does it use AI?<br>• Where is it located? How widespread is it?<br>• Would it interfere with other technology?<br>• Is the data reliable?<br>• Has it been tested? |
| **Governance** | • How is data collected and transferred?<br>• Who owns the data?<br>• Who has access to the data?<br>• Who is using DAS data?<br>• How do we protect or control or manage our personal data?<br>• Who is paying for DAS? The government or private funding?<br>• Who is buying DAS?<br>• Is DAS susceptible to cyber-attacks? Can it be hacked?<br>• On what scale does it operate (by sector, by geography)?<br>• Is it going to be commercialised?<br>• Do we have a say in whether it gets implemented? |
| **Environmental** | • Are there any safety concerns? Is it dangerous?<br>• Is it disruptive to install and maintain?<br>• Will it impact or damage the environment?<br>• Does it require new infrastructure? Especially in rural areas?<br>• Is it temporary or permanent?<br>• What happens when it becomes defunct? Does it get recycled?<br>• How much energy does it consume? |
| **Societal** | • Why DAS and not other existing technology? What does DAS add?<br>• How are the results shared and communicated?<br>• What is DAS to the average person?<br>• Is it good value for money?<br>• What are the risks?<br>• Is it a consumer product? Is it for the layperson?<br>• How would it impact and better my life?<br>• Is it going to make other things more expensive?<br>• Is it government-led or commercially-led? |

# Five research themes

Deliberations during the second and third workshops typically revolved around the five themes: environment, city infrastructure, health and community, surveillance and crime, and governance and costs. In the following we describe and interpret how the participants engaged with each theme.

## Environment

Out of the five themes, the environment generated relatively less discussion than the others. Those who discussed it often saw environmental protection as an uncontroversial public good and were enthusiastic about the potential for DAS to support environmental monitoring and wildlife conservation. One participant was excited by a range of applications, particularly around understanding and protecting wildlife in their local area:

> *"I'm very interested in a place called Campfield in Winchester. It's an old army base that's now home to a lot of endangered animals. Would [DAS] be able to pick up on the number of species and where they're making their nests? What development could take place without disturbing nature?"*

> *"[The] environment was the one that really drew me in the first time round. I'd be very interested to see the scope and the possibilities that the technology offers in terms of assessing, monitoring, and actually planning around wildlife in the countryside."*

Other panellists shared concerns about the possible impacts of DAS on the environment, from digging and placing new cables to the energy demands (and associated costs) of using DAS and storing and analysing the data. They were concerned about *"the environmental detriment that a data centre […] uses to then outweigh environmental damage [compared to] what it can gain".* One participant noted:

> *"I think the most concerning thing is the cost and the environmental impacts […] like a certain amount of data is enough [energy] for 25,000 homes […] that's concerning regardless of what it's used for."*

Not everyone was convinced that DAS would be useful for environmental applications. For instance, one participant with a background in engineering felt it was important to deprioritise the theme as they struggled to see how DAS could be beneficial in this context:

> *"I'd love to see benefits for nature, but if you've got sensing technology through the woods or something, how on earth is that going to [help]? [...] [F]rom my engineering background, I can't understand how that can help the environment... I'm really struggling as to how [DAS] can help nature because nature is going to do its own thing."*

For some participants, however, environmental applications felt more tangible than others, such as health. Others argued that environmental and human health could not be separated, since the environment is fundamental to a healthy planet: *"I initially voted for health, but I think, with the environment, I could just see more of the practical implementation of it."* Others framed the environment as something impossible to avoid:

> *"I think everything will impact the environment no matter which way they go. Hopefully once they decide how to use this, they would have factored in the environment on all aspects, not just to animals, but to us as humans, to the green space, the trees, everything in that regard. So it sort of covers all bases for me."*

The discussions also influenced some people's thinking. One participant explained that the deliberations had *"really cemented my view about focusing on the environment because I feel like we're at a tipping point as humans on our planet."* This was one of several instances where participants explicitly referred to the views and arguments of others when explaining changes in their priorities.

Another argument was that the environment should be prioritised because it would allow DAS to be used for the "greater good" compared to more surveillance, for example: *"doing good was far more important than monitoring people, monitoring footsteps, like a police state type thing, but monitoring the environment, I felt that was overwhelmingly better."* This was echoed by concerns that surveillance and commercial applications would ultimately receive greater attention than environmental uses despite offering less public benefit:

> *"I always think how likely is it that these things will be used for positive reasons that potentially don't have such a financial benefit for large companies and the government, compared to uses that don't help the environment and species and climate change and all of that."*
>
> *"I think infrastructure efforts in that area will get hijacked by the private enterprise and taken off down private routes. I think there's more risk from the surveillance point of view if you go down that route. So for that reason, I think that environment as a priority has less of a risk of that aspect potentially."*

Blurring the lines between infrastructure and environment, one panellist recognised how their priorities would change depending on how the themes would be defined or the context of the issue at hand:

> *"I'd put environment as less important, but I went down the route of floods as infrastructure, not as environment, so in that case, it's the same kind of thing. I'm thinking environment is like wildlife and stuff like that, and I don't see how [DAS] would help with that."*

This difficulty in categorising certain applications was closely related to broader questions about scale. Several participants felt that DAS was most useful for addressing large-scale challenges rather than issues affecting individuals directly: *"It seems like it's only going to be beneficial on large-scale issues. So like earthquakes and floods. I just don't see how it would come down to it being personal."* Similarly, another participant prioritised infrastructure-related applications because they found them easier to imagine in practice:

> *"We're sort of thinking the main priority is large scale disaster prevention, big like infrastructure issues, that sort of thing. Only because it fits in most of that one. If I had my way, it'd be around health and safety, but I don't know how because I can't visualise it and see how it would help it."*

At the same time, discussions prompted some participants to reconsider their initial assumptions about what DAS could achieve. Reflecting on the arguments made by others during the deliberations, one participant explained: *"I'm moving away from city infrastructure towards the nature because of the arguments I've heard [...] There's a lot that can be done for nature [...] So, I think that's really, really doable."*

## City infrastructure

Participants frequently linked DAS to city infrastructure because it was one of the easiest places to imagine concrete, vibration-based applications. This was especially true in the first sessions, where familiar, tangible topics made it straightforward to picture how the technology might work: *"City infrastructure is probably the easiest to understand [...] Identifying pipes underground or monitoring traffic to make sure traffic flow is more effective, that's easier to see how that could work."* They often considered infrastructure *"the most useful and non-invasive" application, including for "urban planning and zoning around airports and other noise polluted areas."*

However, as the Panel progressed and people gained a better understanding of the technology, they tended to downgrade infrastructure: *"City infrastructure was one of my priorities, but I didn't put that one [...] because I think I can understand what it could be used for. But [...] I'm very curious as to how DAS would help with the health category."* Many still voted for it as one of the priorities, driven by the fact that they could understand its potential applications better, despite their interest lying in other areas. Others thought that although it was a feasible area of application, it might not be the domain where the technology could make the most impact: *"Although it's important to know the traffic, is it really that important? […] Maybe it could be used for better things."*

Often, this shift appeared to come through deliberation itself, as participants exchanged views and reconsidered infrastructure in relation to other issues in the greater scheme of things:

> *"I basically put city infrastructure and potholes only because in my road […] there must be about 50 potholes […] But on the grand scheme of things, if you put it next to threat of human life, that's not really that important."*

The discussion around city infrastructure not only revolved around how DAS could be used to help monitor infrastructure, but also on the role of DAS itself as a new form of infrastructure, including questions about maintenance and deployment. Reliability was seen as a concern for the DAS system as a whole, rather than the reliability of individual devices. Participants asked questions such as *"is the technology […] impacted by the age of fibre?"* or *"if we become heavily reliant on this technology […] what happens if those cables do break?"*

Coverage and representativeness also mattered: without a clear map of dark fibre, some struggled to see how city-wide insights would be feasible: *"is there a record of all the dark fibres in the city? Can we see where they all are?"* Others commented on the same issue regarding triangulation: *"I always worry that given that you have to collect and maintain in terabytes of data a second because you have to triangulate to get a location."*

In this context, participants also tended to judge DAS against familiar tools for the same purpose (e.g., traffic apps, CCTV), so "infrastructure" only felt worthwhile when DAS clearly did something current systems cannot. As one participant put it: *"some of those suggestions for what this could be useful [for], we're already doing with other methods anyway. It's duplicating what we're already doing."* The same concerns were echoed across several breakout rooms:

> *"If we've already got something in place that's already doing that […] why do we need something else to replace that and measure vibration if it's going to still alert us?"*
>
> *"There's different apps that you can use as Waze that tells you whether there's traffic, so maybe it could be used for better things."*
>
> *"Isn't this done by cameras already? Do we need extra technology to monitor?"*

Several people concluded that city use needs selective monitoring of exceptions or alerts, rather than continuously monitoring routine activity. This was another reason why they cooled on routine traffic uses while still liking other infrastructure applications. Across different breakout rooms, participants collectively reasoned through the feasibility of different applications, refining and testing ideas against practical concerns rather than simply expressing fixed preferences. In general, some struggled to understand the feasibility of always-on monitoring in an environment as busy as a city: *"But it's difficult to make work on a micro level […] maybe what we should be focusing on is to look at understanding the normal vibrations and identifying exceptions only."*

Inevitably, the discussion bled into other themes (environment, public health, surveillance, and governance) whenever infrastructure touched people and places, with participants stating that "*city infrastructure and community and health are intrinsically linked."* Some offered complex examples that blurred the lines between topics:

> *"If you were able to track heavy vehicles moving along […] and then detecting when they become immediately lighter and use that correlation to say […] a large load of industrial waste*

> *was dumped here. And then you'll be able to prosecute that person, hopefully, and discourage fly tipping."*

In sum, participants initially came up with several ideas of how DAS could be used for city infrastructure, mirroring those promoted by industry, before they had been exposed to the broader possibilities and trade-offs associated with DAS. As sessions progressed, these ideas were frequently deprioritised as deliberation broadened the frame and comparative value, feasibility, and equity came into view. Nevertheless, some participants still tended to prioritise this research area, following the rationale that it was more realistic and practical than others.

## Health and community

Although it was one of the initially prioritised themes, participants often struggled to understand where health "fit". Is the use of DAS technology and data to support understanding of how the environment (air quality) differs between places or how it affects the health of the population? Several participants saw potential value in using DAS to better understand environmental conditions that affect people's wellbeing. One participant reflected on how such information could help people manage existing health conditions:

> *"I can see the potential of that being really helpful for people who do suffer from going out in the elements [...] Not so much for me directly, only because it doesn't impact my circle. My son doesn't have asthma or anything like that."*

Others explained that the discussions had made them think differently about the relationship between environment and health. For one participant, exposure to the air-quality maps that were shown in one of the information sessions changed how they understood the theme:

> *"The health of our environment affects us as humans, how well we breathe, which is why today was so helpful because I hadn't considered that until I saw that map"*

Similarly, another participant reflected on how hearing the discussion broadened their view of the possible applications of DAS:

> *"It would be interesting to see if it can help in healthcare because obviously that's a massive thing that always needs improvement and suggestions and the environment. I didn't think that at all at the start, but after hearing it, I think it definitely would be helpful overall."*

This interest was reflected across several room discussions, not only in relation to air pollution but also noise pollution and the potential to improve the lives of people living in urban areas through a better understanding of environmental exposures. Many participants felt this would be a worthwhile use of the technology:

> *"I also think that city living is probably the most challenging for humans and anything that a new technology can do to make their lives safer, easier, and that's why I've included noise pollution would be a great thing."*

> *"But if that technology can be utilised to, you know, assist people to help people out in who suffer with, from noise pollution acutely, then you know, it can only be a good thing."*

However, others questioned whether DAS would provide information that was not already obvious:

> *"Is some of it not obvious? There's bound to be more noise near an airport, near a road, near a railway. That's not something that we need millions of cables to use DAS data to find that."*

The desire to use the data to help or support people and communities was strong, though there was some disagreement regarding the utility or feasibility of DAS data to inform public-health decisions and priorities. One participant explained that they prioritised health precisely because of its potential to improve people's lives:

> *"I'm very curious as to how DAS would help with the health category. And that's why I've put that on there because [...] currently we're investing so much money into research through various charities, etcetera, Cancer Research, etcetera, to find a cure, whether it's for cancer or dementia, etcetera. And if there's another type of technology that could be used to cure people, then 100% I want that to be made a priority if it means that somebody that's suffering from something can live a bit longer and have a better quality of life."*

At the same time, many participants admitted that health was one of the hardest themes to conceptualise. Several wanted to prioritise it but struggled to see how DAS could contribute in practice:

> *"I don't feel confident with the health bracket because I can't wrap my head around it ... other than potentially finding a route to get an ambulance to the hospital quicker, I cannot wrap my head around what they can do with it."*

Some participants explicitly framed their choices as a trade-off between the desirability of health-related applications and uncertainty about whether DAS could realistically deliver them. As one participant explained, *"I wanted to pick health. I would have picked health if I could see a way that it could be used, but I couldn't see a way."* Similar reflections emerged across rooms: *"I think probably that's why I wouldn't put [health] down because I wouldn't understand how it possibly could be, but maybe it is."*

Overall, there was interest in using DAS data to better understand environmental factors that may affect health. Air and noise pollution were mentioned as key considerations, alongside the importance of using data about places rather than people, which may assuage some concerns about surveillance. There were several people across the sessions who expressed scepticism that the data would be useful to support any health aims or priorities. Health is a topic area where there will be overlap with other broad categories such as infrastructure (inherently linked to air and noise pollution from traffic sources) or environment. Demonstrating how health will be better understood or improved with the DAS data is still very much open for discussion. As one participant summarised:

> *"I can think about the infrastructure more clearly, but then the actual like impacts, like public health, the environment, that's a bit harder to work out what this might be used for."*

## Surveillance and crime

Many panellists saw a direct connection between DAS and the expansion of state surveillance. Discussions often focused on establishing what the technology could feasibly detect and whether it could be used to identify or monitor individuals. Across multiple groups, participants sought clarification on the extent of DAS's sensing capabilities, particularly in relation to listening to conversations and personal tracking: *"Is it a real suggestion that it's so sensitive that it could listen, literally listen to a conversation? Or was that just science fantasy?"*

For some participants, uncertainty about what DAS could detect quickly translated into broader concerns about privacy and personal freedom:

> *"[...] if it can track the way you move. It seems a little bit scary."*

> *"It could track how many times you go to Asda shopping and then they could sell that information to someone. And you wouldn't even need to carry a phone or anything with you. So, it's just a massive invasion of privacy, I reckon. You wouldn't even know about it. You just have to take their word for it."*

Others located these concerns within a wider perception that society is already heavily monitored: *"we have no privacy and this is making it 100% worse, 100% worse."* In contrast, some highlighted opportunities to improve safety, especially for vulnerable groups:

> *"If that's something that we could use for the kids to, say, avoid this road on your walk home."*

> *"Can we do safety alerts for young people, i.e. if a school suddenly got an intruder or someone exposes themselves."*

The link between surveillance and crime was perceived as a somewhat unclear and problematic trade-off: if our rights continue to be violated, can this really be justified by more effective crime control? Participants frequently weighed potential public-safety benefits against concerns about privacy and civil liberties, questioning whether these benefits justified the implications for individual rights:

> *"I just don't know how you would incorporate that with certain rights."*

> *"It's a fine line of getting it right."*

A central image in these reflections was that of a criminal gang gaining access to DAS data to carry out criminal operations. These concerns appeared to draw on broader public awareness of cyberattacks, data breaches, and digital surveillance, with participants often imagining DAS as part of a larger networked ecosystem rather than as a sensing technology in its own right. For example, one participant questioned *"how easy might it be for a criminal gang to put in a fibre to gather intelligence from an area of interest?"*, while others framed the issue in terms of long-term data retention, future analytical capabilities, and the safeguards needed to prevent misuse:

> *"Given that there is a huge amount of data, and no one really knows how to use it, it doesn't take a genius to suggest that if a criminal wanted to save down daily data from my street, he could figure out when everyone goes to work or whatever time they come back. I'm wondering what's being put in place to make sure [...] that doesn't happen."*

> *"Is there anything in place to stop against cyber-attacks that seem to be on the rise?"*

> *"The criminal gang could access the database with lots and lots of data about what noises have been generated in that city. And ultimately they could determine when bank staff have left the premises. So obviously it needs to be super secure."*

These discussions were notable because they often relied on assumptions about the capabilities of DAS that went beyond what participants had learned during the Panel and sometimes connected it to a wider set of digital technologies and risks than those directly associated with distributed sensing. People frequently imagined DAS as part of a highly interconnected digital ecosystem capable of generating detailed intelligence about people's movements and behaviours. In doing so, they often blurred the boundaries between what DAS can currently do, what it might be able to do in the future, and what other surveillance or digital technologies already do. At times, this led to misunderstandings about the technology's capabilities and a conflation of DAS with issues only indirectly related to it: *"What about criminals like paedophiles and dark web, can it detect them?"* The concerns being expressed therefore reflected not only perceptions of DAS itself, but also wider anxieties about cybercrime, data misuse, and digital surveillance.

The evident anxiety surrounding questions of surveillance and crime, however, was accompanied by a certain fatigue with the theme of state surveillance. This took the form of criticism of government and police, who were seen as having already collected all possible data, yet often lacking the resources, capacity, or political will to act upon it.

> *"[...] all the information there, the police don't deal with it [...] this won't change that situation [...] the evidence is there [...] they're doing nothing about it"*

> *"If it's not dealt with by the people who should deal with it and go and arrest someone or investigate something, what's the point, you might as well not have the information."*

Some participants reiterated the need to use the technology for good: *"I think they monitor our traffic, they monitor our phones, they monitor with CCTV, but we should use a new technology to some other benefit."* Similar to discussions in other themes, participants wanted DAS to be used for "good" and often questioned the added value of deploying DAS in areas where existing technologies were already perceived to be working effectively. As one of the facilitators noticed when feeding back breakout discussions to the whole group, "*it was felt that some areas, such as crime and surveillance, already had a lot of technologies in place that were working and therefore implementing DAS in these areas would be wasted in an already technologically saturated market.*"

In other words, some participants saw DAS as contributing to a growing danger of surveillance, while others, especially in London, though not exclusively there, suggested that this technology adds nothing particularly new to existing surveillance systems. This divergence itself points to the uncertain breadth of the concept of surveillance, its different forms and capacities: is DAS yet another innovation in the monitoring of citizens, or is it genuinely a major step towards the erosion of their privacy rights?

Two metaphors kept coming up in the discussion: Orwell's Big Brother and Pandora's box. They were once used in the same sentence, layered on top of one another to underline the point: this technology could give the state still greater surveillance powers, and once embedded, it would become much harder to control. Here, distrust of the state seems to overlap with a kind of technological inevitability in the perception of technology and its development:

> *"It's the ultimate Big Brother. It's like Pandora's box. Once you've opened it, you can't put the lid back on."*

> *"If it can be used to track or influence human behaviour, then I think that then opens a huge Pandora's box about applications in terms of surveillance."*

Interestingly, the panellists did not mention how surveillance technologies can in turn criminalise citizens, especially those of marginalised demographics, by race, class, disability, gender, and sexuality.

Although crime and surveillance generated some of the strongest reactions during the Panel, participants rarely prioritised them as areas for future research. One possible explanation is that participants interpreted "prioritising" a research area as supporting its further development, rather than supporting research into which uses might be acceptable, safeguards that would be required, or how to avoid undesirable outcomes, which was a central component of their discussions. This tension was compounded by a recurring sense of technological inevitability, with some participants viewing increased surveillance as part of a broader trajectory that was already underway. As one of the facilitators noted, *"[the] presentation on surveillance seemed to increase concerns around privacy and data which would be collected by the DAS tech. However, there was not a significant push for this to be a priority insight area when participants reflected at the end."*

## Governance and costs

As responses on other themes have shown, many participants were concerned with ultimate ownership of DAS, and how both commercial and government use of DAS would be appropriately overseen. Some framed this in a more disempowered way, reasoning that the technology is going to evolve regardless and expressing a sense of powerlessness to affect its development:

> *"So just leave it [alone]. It's a race for who can find the best use for it and then they make the profit off it. I'm not talking about individual scientists, but in the grand scheme of things, that's how the roll out will happen"*

Among many participants there was also a concern that DAS would be yet another technology that brings about negative consequences and unforeseen (negative) impacts. Drawing on experiences of other technologies, participants worried that harms may only become apparent after widespread deployment:

> *"My partner's a firefighter and the amount of fires they deal with electric cars is insane [...] I'm scared for the next 10 years for these electric cars and what it's going to bring [...] this type of research could be the same type of thing"*

Others expressed a desire to better understand how the technology might affect them in the longer term: *"I feel like we're sort of in the firing line and I kind of want to know like, how is this going to affect us in the long run?"*

The feeling among many that science is done to them rather than with them pointed to a need for more responsible research and innovation. It also helps explain why participants returned repeatedly to questions of governance, even when discussing other themes. These questions shaped how participants thought about all possible uses of DAS, because they affected whether people saw the technology as worthwhile, safe, or in the public interest.

Another significant theme throughout all discussions, which built upon participants' worries about their agency over technological developments, was the cost of DAS infrastructure, improvements, data storage, and regulation. Even where there was an acknowledgement of potential benefits and efficiencies in DAS, some participants viewed DAS as a potentially superfluous bet in the context of the rising cost of living and services:

> *"Would it be the taxpayer who'd be taking on those costs?"*

> *"What's it going to cost us? Everything else is going up in the world and taxes and whatnot. Is this just going to be another thing that's going to be added to it? What's the potential cost that's going to affect us?"*

For some, this was a question of public spending priorities. They reasoned that if public money were spent on DAS, this could have consequences for other services:

> *"If a huge chunk of taxpayers money is going to go on that I'm assuming it's going to affect other areas of where the funding is currently already going, the taxpayer's money. So how does it affect the overall area, the borough for example in terms of funding, will other services be limited as a result of investing money in this?"*

Some participants judged DAS against existing technologies and questioned whether the costs could be justified if similar forms of monitoring were already available:

> *"The cost of doing this is so huge in terms of the computing power and electricity cost that there's no way that we could justify doing this when we've already got mechanisms that measure things like noise condition."*

This was not necessarily resistance to a new technology, but participants weighing the value of DAS against opportunity costs, existing systems, and the likelihood that costs would fall on consumers or taxpayers. This also influenced how they reached their decisions for prioritisation. As one participant put it:

> *"I put governance and costs because I think the discussion we had beforehand as well was about who puts up the cost for this and everything's getting more expensive. And so normally it falls down to the consumer or the taxpayer to pay for it. We'd rather the money go elsewhere."*

Others were more conditional, suggesting that DAS could be justified if it clearly improved efficiency or reduced costs: *"If it can make it more efficient, more cost effective for everybody, then fine."* However, this conditional support was often accompanied by scepticism about whether promises of future efficiency should be trusted. One participant compared this to claims made about AI, where future improvements are promised while current energy and environmental costs remain unresolved:

> *"I'm just wondering if they're actually going to make it more efficient and cost effective or just go ahead with it and then just say, 'Oh yeah, we're working on trying to make it efficient'. Like AI, for example, although it's a big part of our lives and some can argue it's really helpful, it also consumes a lot of thought and energy and it's really bad for the environment and they're constantly saying, yeah, we're working on making it more efficient, but are they? Seems like they're not."*

Participants recognised that commercial companies might provide funding for DAS products, but reasoned that government regulation would be necessary to reduce perverse outcomes based on profit incentives. They very much saw government as the lesser of two evils in this scenario, and were generally negative and fearful about the concentration of knowledge and power in the hands of large powerful actors:

> *"If this is about influencing the politicians, what's to stop them from using it in a bad way as opposed to a good way?"*

> *"Would there be anything in place for [companies like BP] to have to divulge to make sure that they aren't breaching any kind of privacy"*

Panellists also wondered about the government's readiness to regulate the technology, or its knowledge about what these actors are doing:

> *"At what point is the government at in terms of trying to put something forward to kind of regulate it and, like, make BP accountable for kind of poking up what they're doing? They're just keeping it to themselves and not saying anything about what they're doing necessarily, which means it's hard to then regulate. It would be interesting to know how far along the government are in terms of considering this properly and putting stuff in place [...] if BP are going off on their own, you know, they're kind of leading the way on it if they're already using it, they can kind of be in full control and do whatever they want with it at the moment. Are the government now falling behind and do they need to take it more seriously if they are?"*

These concerns were linked to uncertainty about ownership and control. Participants worried that if DAS were treated as an asset to be bought, sold, or controlled by powerful actors, the public would have limited visibility over who was using it and for what purposes:

> *"At the minute, it kind of sounds like it's in the air and it's like, who's the highest bidder that can potentially take control? [...] We ain't going to know who that is or what their plans are for it, what they're going to use it for [...] So, it's kind of scary that potentially maybe every year it could have a new owner."*

> *"What they could use the technology for, who's going to put constraints on them to make sure they're not using it for bad things?"*

Participants also reflected more familiar critical or cynical concerns about peak industries' strong influence on government, mistrust of political representatives, or general failures of democratic governance to deliver on promises in an efficient manner.

> *"Yeah, because if [the companies have] got money to fund it, they can influence what the government are aware of, and that would influence what decisions and legislation is made on it, because they could say, 'Oh, it's really beneficial for this', to hide all the negative impacts that they could gather."*

> *"I feel like [politicians] have their own ideas, and they just take that forward. I don't think they actually listen to people"*

> *"Yeah, I don't want to be too cynical about it, but I do think that [politicians], yeah, massively, like, use their own agendas."*

> *"There's a lot of red tape [from governments] to grant funding for stuff [...] private companies taking up the technology and the government being late to the party and putting in legislation and stuff in place and using it for the benefits [...] I think [the government will] go, great, but we haven't got the money for it [...] how would you convince the government with all the red tape behind everything to actually do something proactive rather than just waiting for something to happen and then they respond to it. "*

Participants' discussions about implementation reflected sophisticated debates about when and how regulation should occur at different technology readiness levels, and whether general-purpose legislation on data protection or more specific legislation on DAS applications is necessary to achieve desired outcomes. For some, regulatory frameworks already in place provided reassurance about privacy, but they were concerned that existing legislation might leave gaps and called for more proactive regulation:

> *"If we have gone GDPR route, stuff like that, how can we get this information back to us? Because it's ours, isn't it?"*

> *"There's stuff in place actually with GDPR that it protects us to a certain point from that. But then if this wasn't legislated correctly, that could be a backdoor method to get the same data because it's not being legislated to stop you from doing that."*

Others argued that regulation should be in place before wider deployment, rather than emerging only after problems become visible:

> *"I don't think any of this could really be put out until regulations are put in place."*

*"I think it normally takes something to happen for you to do something, it's just reactive. But I guess to be proactive, you'd have to say, this is happening imminently and you've got no control over it, you need to get control over it, otherwise it's going to be a bad day for you."*

*"It all starts with what people are allowed to and not allowed to do, what rules are set [...] the whole crime stays part of it, listening to conversations [...] If it can do that, but you've got something in a legal framework in place and say you're not allowed to do it and you'd be in breach of something, companies would have to back away and not use it."*

# Final voting and research priorities

At the final in-person workshop, participants were asked to move from broad areas of interest towards more specific research priorities for SOUNDSCALE. Before the session, they had been invited to prepare one or two questions about DAS, based on what they had heard in the previous workshops. At the beginning of the day, those who had not prepared were given time to come up with questions or areas of interest they were most interested in exploring. These initial questions were placed under the five research area posters—environment, health and community, city infrastructure, surveillance and crime, governance and costs—or under "something else" if they did not fit within those categories. The full lists of these questions and areas of interest are included in Research priority voting in the Appendix.

The first stage of voting asked participants to identify broad priority areas for SOUNDSCALE to explore. Participants were given two green dots to mark which of the five thematic research areas they thought were important, and two red dots to mark areas they thought were less important. These votes helped capture participants' early priorities at the start of the final workshop, but they were not treated as final decisions. Participants were told that they would continue developing questions and research ideas throughout the day, and that they could reflect on how their views evolved as they took part in further activities. Table 5 shows a summary of this early voting, by location.

**Table 5.** Early voting on broad research areas at the start of Workshop 3.

| Thematic area | Hampshire | | | London | | |
|---|---|---|---|---|---|---|
| | Priority | Not a priority | Net sum | Priority | Not a priority | Net sum |
| City infrastructure | 4 | 32 | -28 | 12 | 4 | 8 |
| Surveillance and crime | 17 | 20 | -3 | 4 | 18 | -14 |
| Governance and costs | 16 | 3 | 13 | 6 | 7 | -1 |
| Health and community | 7 | 13 | -6 | 9 | 11 | -2 |
| Environment | 21 | 6 | 15 | 14 | 2 | 12 |

This early voting showed some important differences between the two locations. In Hampshire, environment and governance and costs received the strongest net support, while city infrastructure received a very high number of "not a priority" votes. In London, environment and city infrastructure received the strongest net support, while surveillance and crime was more clearly deprioritised. These differences should be interpreted cautiously, as participants were voting on broad themes at an early stage of the day, before further discussion, the scenario-building exercise, place-based activities, and final voting.

Before the final workshop, the academic members of the SOUNDSCALE team also carried out an internal ranking exercise, after the five research areas had been defined, to reflect on which areas they personally thought should be prioritised. This was never shown to participants and did not determine the workshop activities, but it provides a useful point of comparison. The original exercise used 1 as the highest priority and 5 as the lowest. For ease of interpretation, Table 6 presents the individualised anonymised scores, where 1 indicates lowest priority and 5 highest priority, and the totals therefore represent the overall priority score, with highest numbers indicating overall stronger support.[16]

**Table 6.** SOUNDSCALE team internal ranking of research areas; higher scores indicate stronger priority.

| Team member | Health and community | Environment | Surveillance and crime | Governance and costs | City infrastructure |
|---|---|---|---|---|---|
| **1** | 5 | 4 | 3 | 2 | 1 |
| **2** | 4 | 5 | 3 | 1 | 2 |
| **3** | 2 | 3 | 5 | 4 | 1 |
| **4** | 3 | 2 | 1 | 5 | 4 |
| **5** | 1 | 5 | 4 | 2 | 3 |
| **6** | 1 | 2 | 3 | 5 | 4 |
| **7** | 2 | 4 | 1 | 3 | 5 |
| **Total** | 18 | 25 | 20 | 22 | 20 |

The internal ranking showed variation within the research team itself. When the rankings were combined, environment emerged as the strongest internal preference, followed by governance and costs and very closely by surveillance and crime and city infrastructure, with health and community receiving the lowest overall priority. The internal variation shows that there was no single pre-existing "team view" for the Panel to confirm. It also highlights the value of asking citizens to deliberate on research priorities: participants approached the same themes through their own concerns about public value, cost, regulation, privacy, environmental protection, feasibility, and local relevance.

The "selected priorities" from participants were not chosen from the initial list. Rather, they developed their priorities further over the course of the day. Panellists generated new questions and ideas through the science-fiction world-building exercise, discussions with SOUNDSCALE researchers, and the place-based walk or map activity. After the world-building exercise and researcher discussions, participants placed new research ideas on the relevant theme posters and voted again, this time using one green dot and one red dot to identify individual ideas they wanted to put forward or reject. All post-it notes that received at least a green dot were then collected under a "selected priorities" poster, with facilitators grouping or merging similar questions or ideas where appropriate.

At the end of the session, panellists were asked to vote on their final priorities. Each participant was given one final sticker to place on the selected priority they most wanted SOUNDSCALE researchers to explore. They were then asked to discuss their choices, reflect on whether their priorities had changed throughout the day, and write down their final choice and reasoning in their workbooks. Table 7 and Table 8 summarise the selected priorities that received at least one final starred vote:

[16] Only two members did not provide their priorities on time.

**Table 7. Selected priorities receiving final starred votes at the end of Workshop 3, Hampshire. Questions are reproduced as written by participants; grammatical and syntactic features have been retained.**

| Area and questions (Hampshire) | Starred |
|---|---|
| **Environment** | |
| Having a planet to live in is more important. | 1 |
| We should prioritise our planet and anything that could help with the climate crisis. | 8 |
| Could there be an international DAS for the environment collaboration? | 1 |
| **Health and community** | |
| Are there any health uses? | 3 |
| **Surveillance and crime** | |
| Can we track an individual across different cables at depths amongst all the background noise? | 1 |
| How will this affect us with the surveillance and our privacy that we need? | 3 |
| **Governance and costs** | |
| How can DAS be regulated by the government so that private companies can't monopolise its use? | 5 |
| Have the cost implications and energy usage for the data centres been calculated vs the value of the data gained/potential usages? Is it worth it? | 1 |
| **Something else** | |
| How is DAS (and our potential reliance on it) affected if the fibre optic cables break? | 1 |
| How fast can the data be processed? | 2 |

**Table 8. Selected priorities receiving final starred votes at the end of Workshop 3, London. Questions are reproduced as written by participants; grammatical and syntactic features have been retained.**

| Area and questions (London) | Starred |
|---|---|
| **Governance and costs** | |
| How can we regulate the people/users using the technology – Licensing? | 2 |
| Who is currently interested, investing and researching the DAS technology? | 1 |
| How do we monitor and regulate through its evolution? | 5 |
| Cost vs benefit of the project [of DAS deployment or development]. | 5 |
| **Something else** | |
| What are the disadvantages of not using DAS? | 5 |
| Why?[17] [sic] | 2 |
| If [DAS] can add on tech [sic] rather than just pick up sound? | 1 |
| Can the cable be used for multiple areas to address? Regarding different locations and every multiple [sic] research area? | 1 |

The final starred votes showed continued interest in governance, regulation, cost, environmental questions, and the wider justification for DAS research. In sum, governance and costs received the highest number of final starred votes. Participants prioritised questions about how DAS could be regulated, who would own or control the data, how the technology could be monitored as it evolves, whether licensing might be needed, and whether the costs and benefits of the development of DAS justified further research. In Hampshire, a question about regulating DAS so that private companies could not monopolise its use received five final starred votes. In London, questions about monitoring and regulating DAS through its evolution, and about the cost-benefit of DAS, each received five final starred votes.

Environmental priorities were strongly represented in Hampshire.[18] The highest-scoring individual priority was the statement: "*We should prioritise our planet and anything that could help with the climate crisis*", which received eight final starred votes. Other environmental priorities included questions about international collaboration, environmental impacts, wildlife habitats, data centres, and whether DAS could contribute to understanding or reducing pollution.

The final voting showed that some participants wanted SOUNDSCALE to consider broader questions about whether DAS should be pursued at all, and what might be lost if it is not. In London, "*What are the disadvantages of not using DAS?*" received five final starred votes, suggesting that some participants were thinking not only about risks of deployment, but about the possible consequences of not developing or researching the technology. Other final priorities included questions about why DAS was needed, how fast data could be processed, what would happen if fibre-optic cables broke, and whether DAS could support multiple areas of research across different locations.

The chosen priorities offer a simple way of seeing which ideas participants wanted to emphasise at the end of the day, but they do not capture the full reasoning behind those choices. As seen throughout the Panel, people's priorities shifted as they learned more about DAS, heard from others, discussed trade-offs, and reflected on the technology in relation to place. Voting also has well-known limitations: results can be shaped by how ideas are grouped, how questions are worded, and how participants understand what it means to mark something as a "priority". For this reason, the final priorities are best understood alongside the full range of material generated through the Panel, especially the deliberations, which provide a much richer account of participants' reasoning, uncertainty, ambivalence, and changing priorities. The discussions, questions, and post-it notes provide a richer account of what participants thought mattered, where they were uncertain, and why some themes became more or less important as the deliberations developed. In this report, we place emphasis not only on what participants voted for, but on how they reasoned, what trade-offs they considered, and how their views changed through discussion.

[17] We interpret this statement as broadly saying, "why should DAS be deployed/investigated?".
[18] Whereas in London, the general thematic area of "environment" scored highly, but at the end of the day participants did not vote for individual priorities in that area.

# Reflections on the Panel process

One of the biggest surprises of the Citizens' Panel was the exceptionally high level of attendance and retention throughout the process. We over-recruited to ensure a minimum of 40 participants, but all 50 initially recruited panellists attended the first two workshops, and only two dropped out of the third workshop in London due to calendar conflicts. A total of 48 participants completed all three workshops, which is 20% greater than the target sample size of 40.

This high level of participation is likely attributable to a combination of factors. These include an interest and curiosity about new technologies; the fact that the project was framed as an independent academic initiative rather than being led by industry or government; the use of cash incentives rather than vouchers; the effectiveness of recruitment through established local networks in London and Southampton; and the high amount of dialogue between the research team, who were present in all sessions, answering questions and providing further information. Recruiters drew on local relationships and advertised in relevant community forums, which also likely contributed to high turnout and sustained engagement across multiple sessions.

The additional recruits tended to be from higher social grades, however, since some of the lower social grade quotas were not met (see Quotas in the Appendix). This was partly due to a smaller proportion of lower decile postcodes available close to the NDFF cable in Hampshire. Many fibre-optic cables already serve more affluent than poorer areas, which was something the research team had identified as a potential issue for DAS technologies at the outset of the research. The recruitment problem also reflects wider challenges in engaging people from lower social grades in research, which may have been heightened by the technical nature of the topic. Although a high cash incentive was offered to overcome some of these barriers and good representation across educational attainment quotas was nevertheless achieved, this pattern suggests that future projects may need to give additional consideration to fostering closer ties with local community groups and charities, and to how emerging technologies and academic research are framed and communicated in order to better engage underrepresented groups.

## Understanding unfamiliar emerging technologies

In developing the workshops, we were concerned that the participants would find it difficult to understand distributed acoustic sensing. We were also cautious about priming them by giving examples of specific DAS applications which could bias the panellists towards specific ideas. In the end, we introduced DAS by briefly explaining its working principles and how the technology is used in maritime settings. This enabled participants to come up with their own ideas for how DAS might be used in cities. More in-depth explanations of potential use cases of DAS in cities followed in the second workshop, drawing on the academic literature and industry interests. The aim was not to withhold these ideas from participants, but to introduce them at a later stage, after they had already had time to discuss the topic and form their initial reactions.

Overall, we believe that this phased approach—introducing the technology through a different setting, maritime environments, before asking participants to consider how it could apply to cities—was appropriate for this Panel. It helped avoid participants becoming fixed on specific urban applications too early, and the evidence shows that some changed their minds as the deliberations progressed (see Five research themes). One disadvantage, however, was that some were initially hesitant to propose ideas in case they were incorrect or came across as ignorant. Even so, the citizens came up with mostly relevant themes and scopes of application for DAS, many of which matched existing directions of research and potential deployment. On the other hand, a couple of panellists claimed that they had grasped DAS very quickly and made overconfident statements that contained factual inaccuracies which went by unchecked. This meant that those groups deliberated on somewhat inaccurate information, but the disruption overall was minimal. This highlights the complexities of deliberating on emerging technologies that are still not part of people's technological imaginaries, where participants may reason thoughtfully while still working with partial or uncertain understandings of the technology.

Some of the misunderstandings arose because of the connection between DAS and fibre-optic cables. This was confusing for some who were led to believe that the purpose of DAS was to provide faster internet or identify connectivity issues. Others, for whom health was of key importance, wanted DAS to be used for medical applications, or worried it might cause health issues:

> *"There must be something they can use it for in the medical world."*

> *"Identifying high areas of pollen if a child has asthma."*

> *"Maybe health risks? Because when 5G came about, people were saying it causes cancer."*

We also saw how certain nuances around emerging technology were missed. A couple of participants noted that they had done their own research on DAS coming into the second workshop, describing examples of DAS being used over 10 or more years ago.[19] This revealed some confusion around why DAS had been framed as emerging.[20] It also reflects a misunderstanding of the pace of development of technologies and highlights further challenges in trying to explain DAS to citizens:

> *"I thought it was quite a new thing, but it's not actually new. I read up about it, I didn't realise it had been going for the past 20 years."*

> *"On the information we received from yourselves, it says it's a new technology. But when I researched about it, it says it's been going for about 10 years. I'd be quite interested in what's actually new, and if they've been using it for the last 10 years, how can it be new?"*

In reflecting on the role of the different workshop activities to elucidate DAS and the SOUNDSCALE project, the 'world café' (see Workshop 2) was especially helpful for some, as well as a brochure created and shared by the research team. The brochure explained the working principles of the technology and aims of the project, so the panellists could read at their own pace. The more that the participants understood DAS, the more they were able to enjoy the workshops:

> *"I understand it better now. I think before I was quite confused about the ways in which it could be used. It's making a lot more sense in my head and I'm finding it a lot more interesting now as well."*

> *"Tonight has made me understand it a hell of a lot more than I did last time. I can now envisage how it can be used in different ways. So, tonight's been a real sort of eye opener for me to look at it in a different way."*

> *"The documents that we were supplied before this briefing were really useful as well, just to get a bit more information."*

This reveals an additional challenge in deliberating on largely unfamiliar emerging technologies. Participants needed time and different ways of learning (e.g. through presentations, text, discussion, and Q&A) to develop a basic understanding of DAS, a process that unfolded unevenly across the Panel and took at least two sessions. During this phase, discussions were often marked by heightened uncertainty, discomfort, and scepticism. This was not necessarily a failing of the process. Learning about unfamiliar technologies often takes time and reflection, but is worthwhile, as it is challenging and delivers novel insights. Some discomfort is to be expected when participants are being asked to engage with concepts that are new to them. In this sense, the Panel confirmed the value of a deliberative approach: participants were able to reach a basic understanding of DAS through two online sessions, supported by a clear and realistic learning goal, repeated opportunities for questions, and constant dialogue with researchers, which we believe was a very effective learning tool.

[19] Distributed accosting sensing and other similar sensing technologies were initially employed by the military in the 1970s for intrusion detection, and later by the oil and gas industry in the 2000s.
[20] Emerging technologies can be used for years in niche settings while they are being developed before mass adoption.

On the other hand, the phased introduction of information, while successful in helping participants reach a basic understanding of DAS without becoming fixed on specific urban applications too early, created some challenges. Because DAS was introduced in two stages with a two-week gap, some left the first workshop without a full understanding of the technology or its possible urban applications. The gap between sessions may have allowed some uncertainty, scepticism, or partial understandings to develop further before they could be addressed. For practitioners of deliberation on science and technology, one practical learning is that combining a clear learning goal with interactive dialogue and periods of reflection can work well, but the timing of those stages matters. Where information is deliberately split, it may be better to do this within the same session, or with a shorter gap between sessions, so that participants have time to form their own initial reactions without being left for too long with limited or incomplete information. Practitioners should also consider sharing information in different formats, to account for the multiple ways in which people learn and comprehend.

# Understanding their role in the Panel

While some uncertainty was expected around understanding a novel technology such as DAS, a more persistent challenge concerned participants being initially confused about their role in the Citizens' Panel, its purpose, and how their contributions would be meaningful. This was particularly evident in the first workshop, where participants questioned the value of their input as non-technical experts and struggled to locate themselves within this deliberative process:

> *"I can't really understand how we can shape [the technology]. I don't really understand what you need from us."*

> *"I'm not sure how much value we can all add."*

Despite this uncertainty, participants were proactively and thoughtfully engaging with the topic as they became more comfortable with it. This was evident from the wide range of questions asked about the technology and its implications, as well as those described in the Initial reactions to DAS section.

The team began the Panel by providing information about the project, the goals, and the role of the participants in shaping the research agenda. However, in hindsight, our explanations did not provide enough context about how academic research is carried out, how research agendas are developed, and how this differs from other types of research, such as market research or consultation on an already-designed product. More explicit context about the research process, the different actors involved, and the role of universities in the science ecosystem may have helped participants better understand the purpose of their involvement and why their views were being sought at this early stage. This confusion was not only about whether they had enough technical knowledge to contribute, but also about what kind of influence the Panel could realistically have: *"I don't know how* *we would get [politicians] on board with what we're saying."* Some participants seemed to interpret the Panel as a consultation on the future deployment of DAS, rather than as a process for informing research priorities of an academic project at an early stage:

> *"Who ultimately gets the say for what it will be used for? I'm going to guess that we're not going to be the deciding factor of what it gets used for, is that why you're researching? It's going to happen, but it's still very undecided on what it's going to be used for?"*

This uncertainty was also linked to wider unfamiliarity with how academic research agendas are shaped. The team tried to communicate the general unknowns of the research process transparently, for instance, when they were unsure about open research questions. For some participants, this openness about uncertainty did not always build confidence: *"I'm finding that they're all a bit vague. They're not giving me much confidence, you know, that they're not sure and everything's so unknown, how can we have any confidence in any of it?"*

Some of this confusion was born out of a genuine shock at the novelty of asking lay people early on in research to engage and influence the research process. In a small number of cases, participants expressed scepticism and believed the team had a hidden agenda. Some of these concerns included assumptions that the project aimed to profit from their ideas on how DAS could be used, that the research was guided by government interests, or that the Panel was a box-ticking exercise from industry developers:

> "*I gather that ultimately this is a government financed project and the government has ideas. They've already got an idea of what they want to use it for, and they just need to justify how they're going to use it because it is an arm of the government. I don't know which arm of the government. It could be security."*

As described in Understanding unfamiliar emerging technologies, this scepticism may have been shaped by the phased approach to introducing the technology. The Panel initially presented DAS through maritime examples to avoid priming participants with specific urban use cases. While this supported more open-ended discussion, it also meant that some participants initially felt they did not have enough information about the technology, its possible applications, or the purpose of our research.

As well as wondering 'why me' in the context of being a lay person, they also questioned the geography of the Panel. One participant realised that they had all been recruited from unusually specific locations, and since we had not yet explained our place-based approach to the Panel (where we recruited using postcodes situated close to the NDFF cable), it aroused suspicion and led some to believe that there was an undisclosed agenda.

Attempts were made to address these concerns by the time of the second workshop. In between both sessions, participants were sent a two-page brochure that provided a clearer explanation of DAS, allowed time for independent reflection, and more explicitly positioned the research project as independent from commercial or government interests. In addition, the second workshop included a clearer explanation of the NDFF and the project's place-based research design, which helped clarify why participants had been recruited from specific locations and reduced some of the earlier scepticism.

These efforts had practical implications for the design of deliberative processes. When phased information provision is used for emerging technologies, sufficient continuity between sessions is critical. Extended gaps between workshops can heighten uncertainty or allow partial understandings to harden before they can be addressed. Alternatively, very explicit communication about why information is provided in stages, what participants can expect at each stage, and what kind of influence they have may help mitigate such assumptions. Note that more detailed explanations will require additional time and can increase cognitive demands on participants, so this needs to be designed carefully into the process.

The panellists' reactions were not universal, and several participants, especially after the second workshop, quickly grasped the purpose of the Citizens' Panel and were enthusiastic about the opportunity to engage at an early stage and provide their views in the development of an emerging technology: *"I get why they want us here [...] There are concerns maybe about privacy, maybe about the limits of its use. I get where the research is going."* For these panellists, the value of the process lay precisely in being consulted before concrete applications or deployments were already decided. Several contrasted this experience with more familiar forms of public or stakeholder engagement, where individuals are typically asked to respond to a fully formed product or policy rather than to shape the direction of research itself:

> *"I think it's fantastic, starting at this point asking the questions now is unprecedented really. I haven't seen that done and so that's really promising […] In my experience, they hold a research group and hope to persuade you to believe what they've already done, and I've been in a few over the years. So, I do think it's really innovative to do it this way."*

> *"One of the things I find fascinating about it is usually when you do these things, you get given a product and you have to make an opinion of it. It's very rarely something new and it's very rarely done before they've started using it generally. So that's quite fascinating as well."*

These participants articulated a clear discernment of their role not as technical experts, but as contributors of values, concerns, and perspectives that might otherwise be absent from research on emerging technologies:

> *"[We are] here to give our views, not to be experts in the technology but to shape the direction of travel on the research going forward […] [There's] no such thing as a silly question."*

However, this level of understanding was not necessarily shared by everyone, pointing to something broader. Even when additional explanations were provided about the position of the project and the role of the researchers, some participants continued to interpret the Panel through wider scepticism about government funding, industry, and technological development.

> *"I think it's slightly pessimistic, but only from the perspective of why do we already do this? Why can we not focus on things that we're already doing and improve those things? Why do we have to now spend a proportion of money investigating, researching? Finding out if this works, there's a lot of time and resources, why can we not think about the technologies or the infrastructures we already have, and in the selecting priority areas that people have identified."*

> *"Are they going to let us know if there are any dangers with it?"*

This mismatch highlights some of the tensions between the realities of early-stage research on emerging technologies and public expectations of technological development. From the research team's perspective, the project was motivated by uncertainty: DAS was treated as an open-ended emerging technology whose future applications, implications, and desirability were precisely what the Panel was intended to interrogate. For some participants, however, this same uncertainty conflicted with assumptions about how research and technological development usually work. As a result, questions surfaced repeatedly about whether the research team already knew how the technology should be used, even when the Panel's purpose was framed as exploratory and as offering an external guide to research on the technology.

Thus, explaining aims and processes is important, but not always sufficient to overcome deeper assumptions about how science and innovation operate, particularly in the context of emerging technologies with which participants have little prior familiarity. Some scepticism ought to be expected, celebrated, and valued; it shows that we had created a safe space where participants felt able to challenge the process, and that the Panel included a range of attitudes towards trust and technology. We will also re-engage with the Panel as our research develops so that we can continue to build trust and they can continue to shape and inform our work. Artistic and community research within SOUNDSCALE provides another way of exploring how people relate to public infrastructure and emerging technologies.

# Framing and designing the Panel

The goal of the Citizens' Panel was to inform research priorities about DAS for the SOUNDSCALE project; it was not to decide DAS deployment or produce a representative account of public opinion. To enable this, we designed the Panel in such a way that the citizens would learn about DAS and explore its implications in different contexts so that they could deliberate and make informed choices. This type of process was appropriate because SOUNDSCALE aimed to involve citizens early in the research process, before the applications, priorities, and implications of the technology have been set. Rather than asking for immediate reactions to a finished technology, the Citizens' Panel created space for participants to learn about DAS, explore uncertainties and trade-offs, and use their own lived experience, values, concerns, and local knowledge to inform the direction of the research.

In this sense, the Panel worked well. Participants asked critical questions, exchanged views, reflected on the information presented to them, revised their priorities over time, and identified research directions that were relevant to the project. The overall quality of the Panel was felt to be high by researchers and facilitators with many years of experience in deliberative engagement. Participants came up with relevant concerns and research directions, while also demonstrating the reasoning, uncertainty, and changes of view that are central to deliberative processes.

At the same time, the design involved some important trade-offs. The Panel was not designed around a settled policy choice, or around presenting participants with clearly defined "for" and "against" positions on DAS. Given the early stage of the technology, many of the possible applications, risks, benefits, and governance arrangements were still uncertain. While the research team sought to be balanced in how it presented possibilities, uncertainties, and risks, we did not purposely select opposing pro- and anti-DAS perspectives. This was because the purpose of the Panel was to inform an early-stage research agenda within the scope of SOUNDSCALE, rather than to produce a general account of public opinion for or against the technology.

This points to a known limitation of using Citizens' Panels to deliberate on highly unfamiliar and early-stage technologies. Panels can create time and space for citizens to learn, ask questions, exchange views, and reason together, but they cannot fully resolve wider issues of trust, unequal expertise, or uncertainty about how a technology may eventually be developed and used. We sought to address this by dedicating sufficient time for both learning and discussion, across several sessions and a mixture of online and in-person workshops, repeated opportunities for questions, direct dialogue with researchers, phased information provision, and opportunities for participants to revisit their priorities as their understanding evolved. The process also showed, however, that more explicit framing is needed around what it means to shape research, what kind of influence participants can realistically have, and what sits outside the scope of the Panel. The artistic research and later community engagement activities in SOUNDSCALE are therefore necessary complements to the Panel, as they allow some of these questions to be explored through more situated, open-ended, and public-facing forms of engagement.

In hindsight, our focus on explaining DAS technology meant that we did not provide enough context about how academic research is carried out, how it differs from other types of research, such as market research, and how this process was specifically about informing the research priorities of the project. We also could have explained the research and development (R&D) process and the different actors involved, and the role of universities in the science ecosystem. Although difficult to do for emerging technologies, we also could have given the Panel a say in who we should have invited to the workshops, a mechanism often used in other mini-publics to help build trust. We think these changes may have helped the panellists to reach a clearer understanding of our agenda, and reduced their scepticism about the project and their confusion in taking part (see Understanding their role in the Panel).

In this regard, the science-fiction aspect of the methodology was particularly useful as the participants reported that it gave them useful tools of literacy and imagination to be able to have a conversation about the emerging technology of DAS. In this setting, using fiction-making and world-building helped mitigate hierarchies that are inherently present in scientific research where scientists can often be presented as holders of relevant expertise and participants as holders of "opinion" or "input", and participants may therefore not feel empowered to fully participate. Instead, the world-building exercise allowed the participants to engage more freely in discussion, and to explore in greater detail how a particular societal, political, technological or cultural "future" would change given the presence of this emerging technology. The participants noted that they enjoyed the exercise and that it created a more accessible space for them to engage. Qualitative science-fiction methodologies which are sensitive to participants' positionality, their specific socio-economic struggles and their experiences, allow the fictionalised methods to empower them and to create conditions for interest, literacy-building and capacity to mitigate hierarchical settings in research.

The use of the term "research priorities" or "priority research areas" throughout the Panel added further confusion. Some participants felt unsure about what these meant and what they were supposed to contribute, or whether they had enough expertise to make a useful judgement. These challenges were anticipated to some extent and were the result of trade-offs in asking citizens to engage with an emerging technology before its applications were fully settled. In designing the Panel, we had discussed whether to include a more explicit learning component on how academic research works, or what makes for a good research question. We ultimately decided not to do this, partly to avoid overloading participants and partly because we did not want to narrow their contributions into academic categories too early. Instead, the intention was that the research team would listen to what participants cared about and then translate those concerns into research priorities.

We also found that people prioritised based on very different rationales. Some prioritised according to what was most important to them personally, others according to what they felt the technology was capable of, or according to what they thought would be the most beneficial to people and planet. This is not necessarily a problem with the process, but it shows the limits of interpreting votes alone. Voting provided a useful indication of priorities, but the deliberations were essential for understanding why participants voted as they did, what trade-offs they were making, and how their views changed over time. These differences are evident when comparing discussions reflected in The Panel deliberations and the priorities reported in Final voting and research priorities.

Differing rationales were particularly visible, for instance, in relation to 'surveillance and crime' and environment. While the Panel expressed serious concern about DAS being taken advantage of by the military, companies, governments, or criminals, 'surveillance and crime' received high numbers of votes for 'not a priority' at the beginning of the final workshop. This might have meant that the Panel interpreted 'priority research areas' as meaning ways in which DAS should be researched for 'good' purposes or applications, without wanting to prioritise research based on the prevention of DAS being used for 'bad'. Environment received strong early support in both locations, and remained a major final priority in Hampshire, but this did not translate into final starred votes in London.

These reflections reveal the nuance involved with using voting systems and taking those results at face value, compared to the complexity of the Panel's reasoning and decision making as demonstrated by the deliberative process. This is perhaps exacerbated when the focus is on an emerging technology of which public awareness is very low and for which it remains unclear who exactly is using this technology and for what. For this reason, the final priorities should not be treated as fixed or written in stone. They offered a simple way to capture impressions at different points in the process, but they need to be read alongside the wider deliberations, where participants exchanged views, tested ideas, shifted priorities, and explained the reasoning behind their choices.

# Acknowledgements

Immense thanks are due to all the Panel participants, without whom this research would not have been possible. Thank you also to the lay experts, Paul and Shanmukh, venue hosts, caterers, and local recruiters from Criteria Fieldwork, who all helped to ensure the smooth delivery of the Citizens' Panel and created a warm and welcoming atmosphere throughout the workshops. Special thanks go to Ciaran Cummins for his earlier work on the Panel design; to facilitators and support staff, Charlotte Bowen, Yasmin Begum, Alessandra Radicati, Claudia Nashef, John Evans, Ellie Mendez Sayer, Claire Pearce, Ella Halepota, Lila Ovington, and Harry Pearse; and to Luke Huckle for filming and producing the Panel video. Last but not least, thank you to our advisory board for their invaluable feedback on the Panel design.

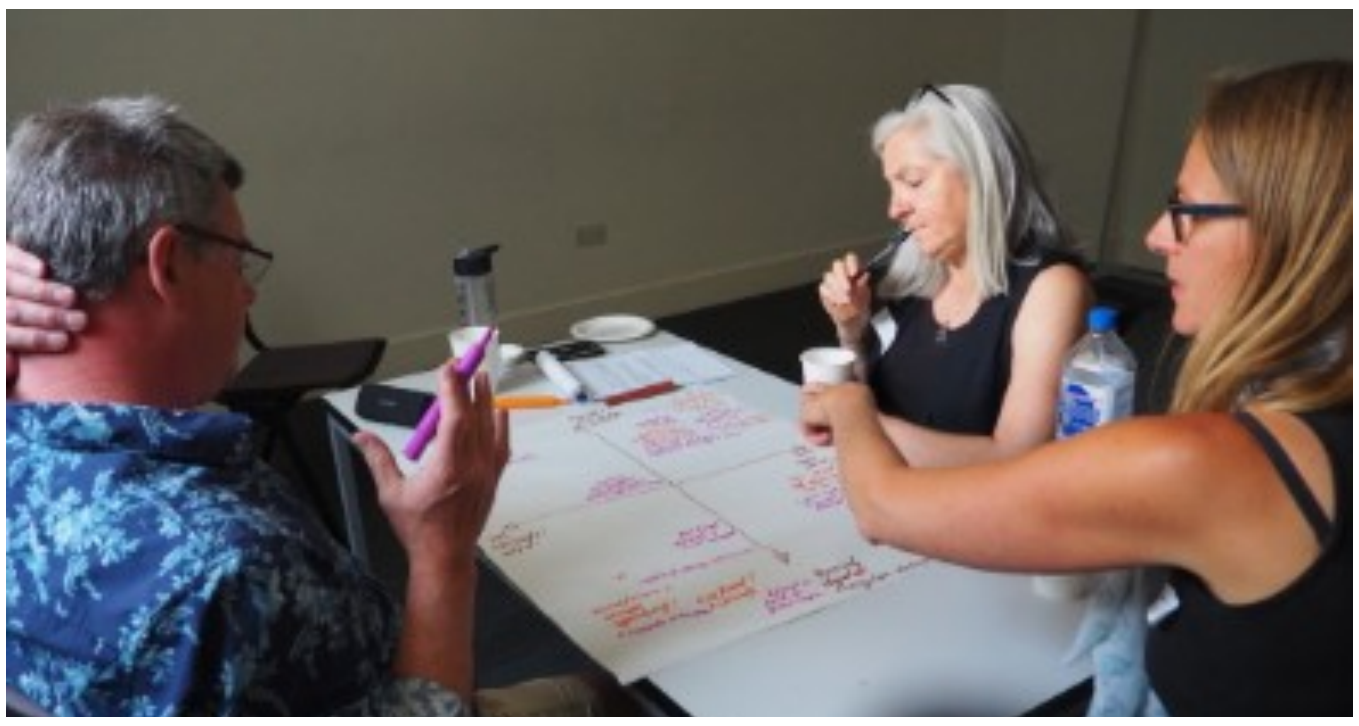

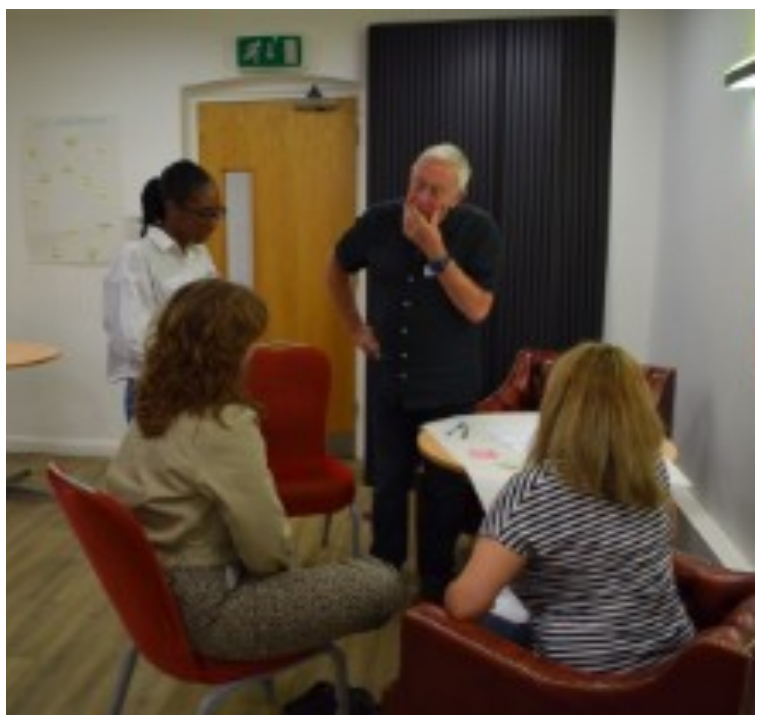

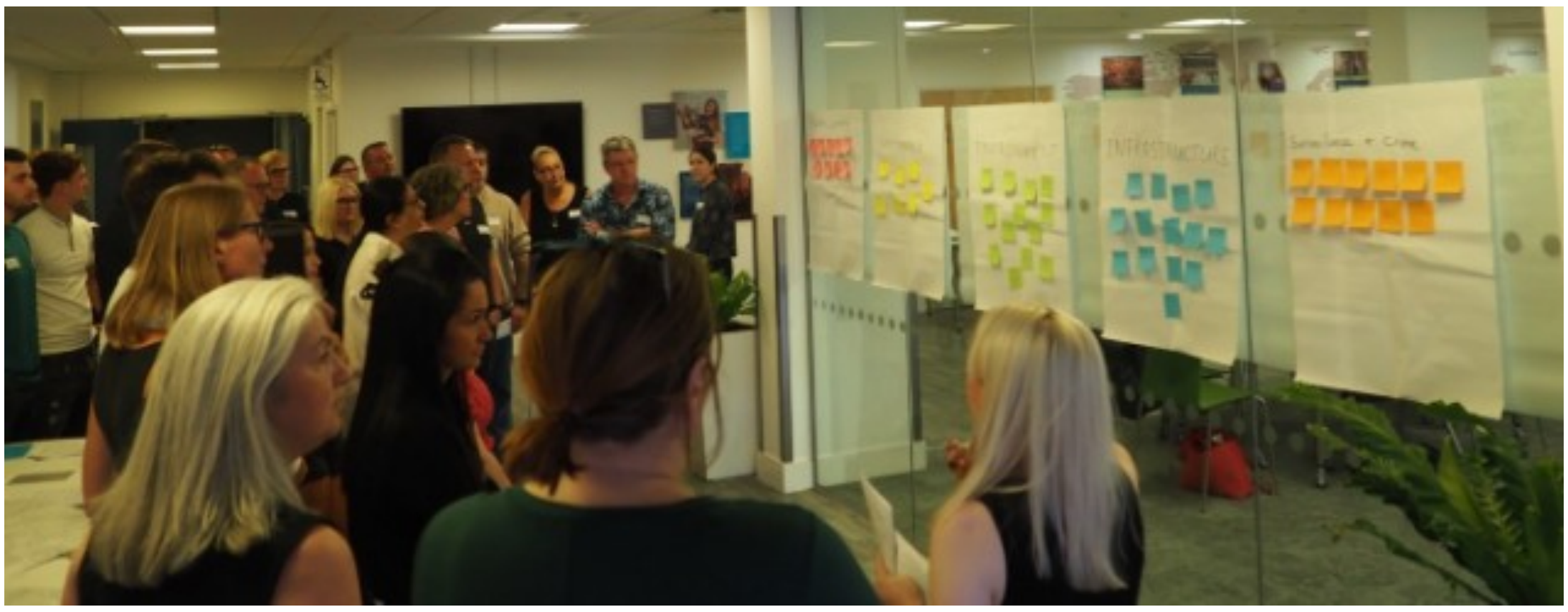

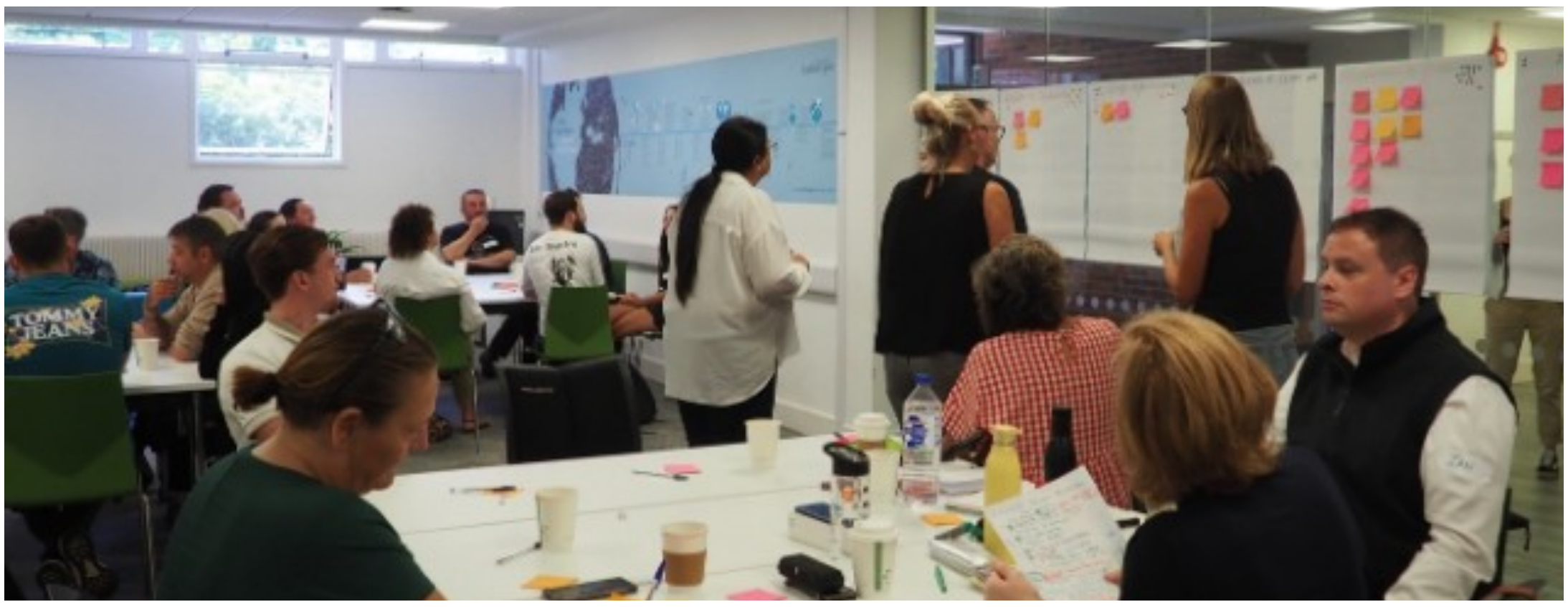

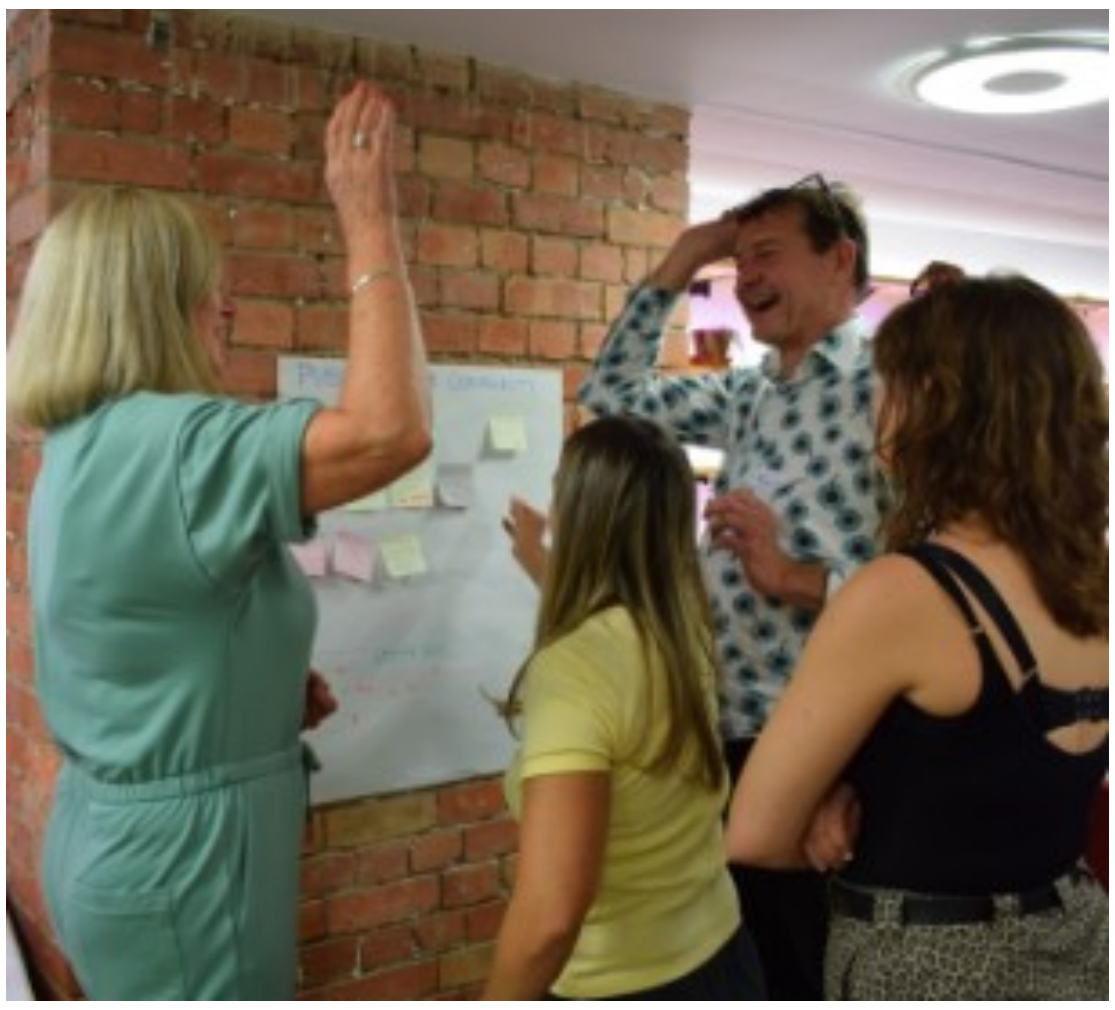

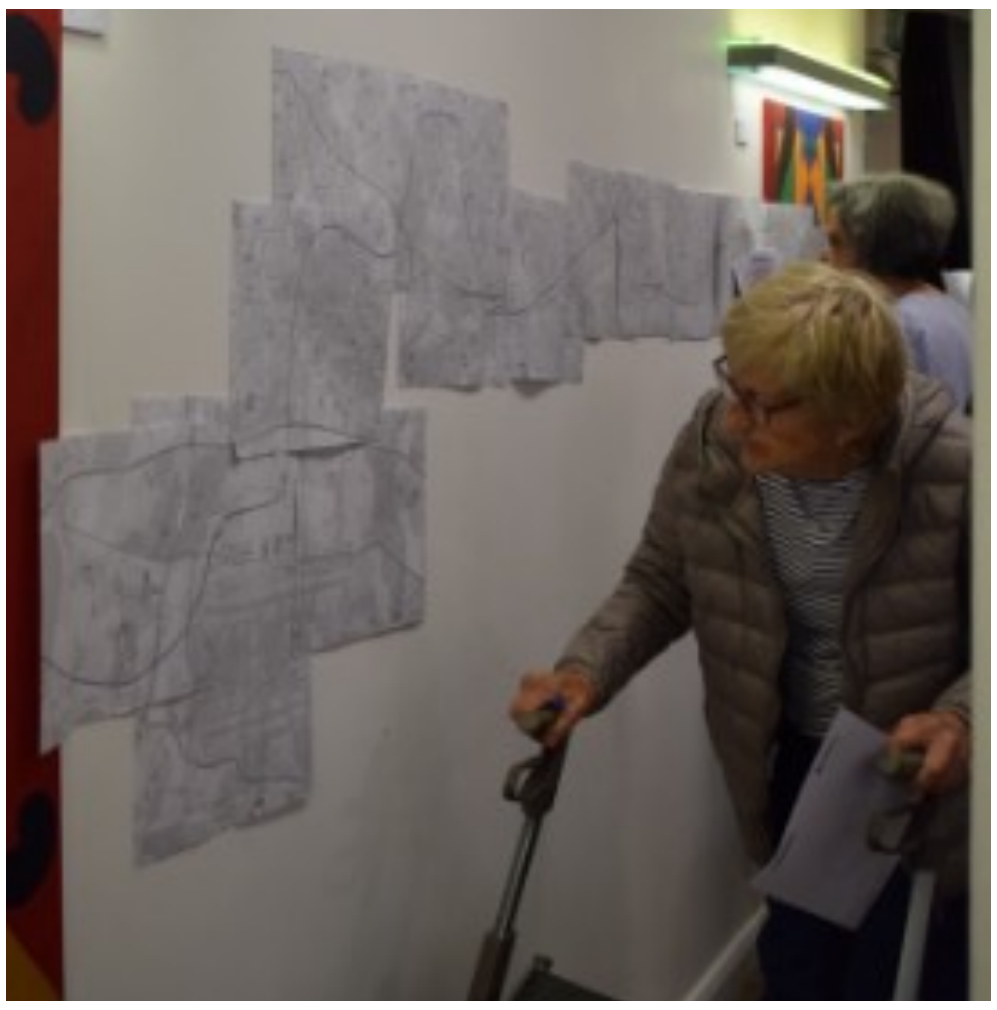

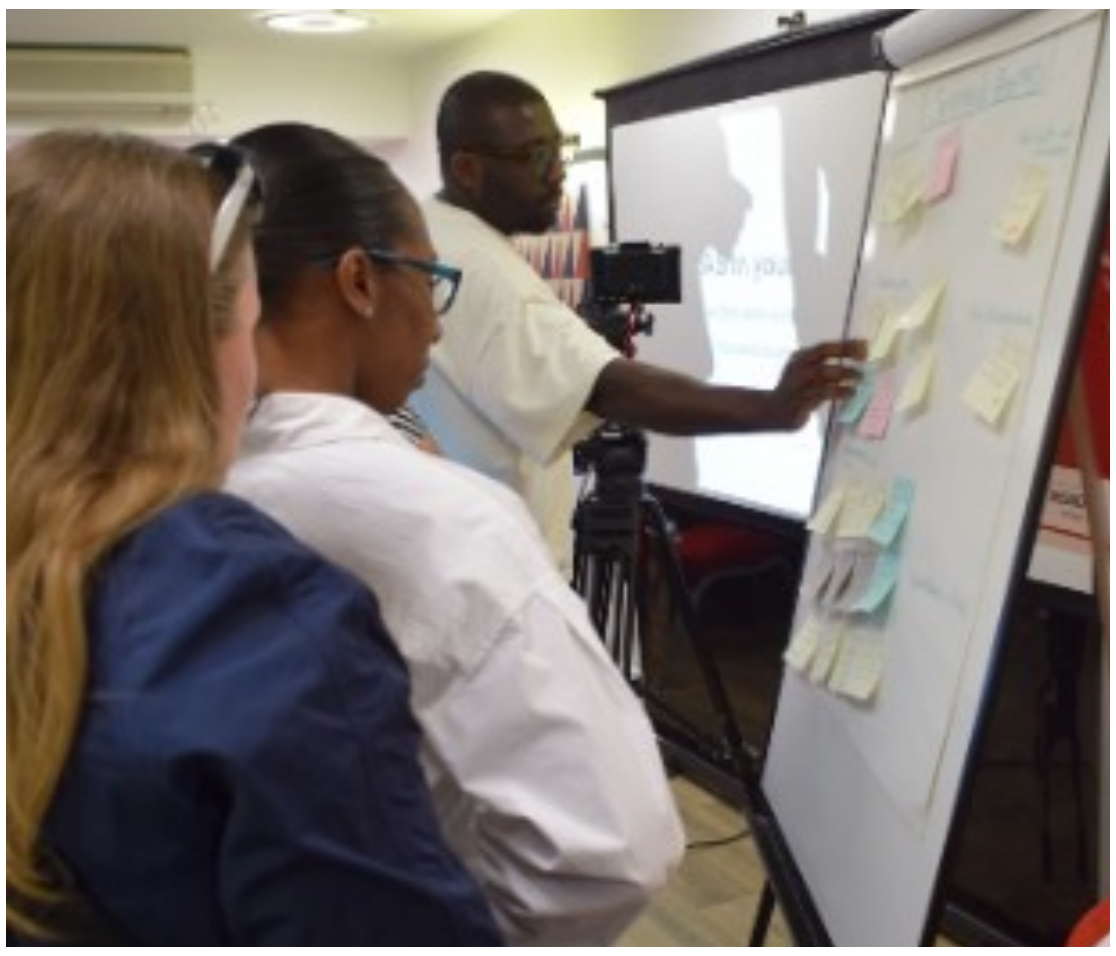

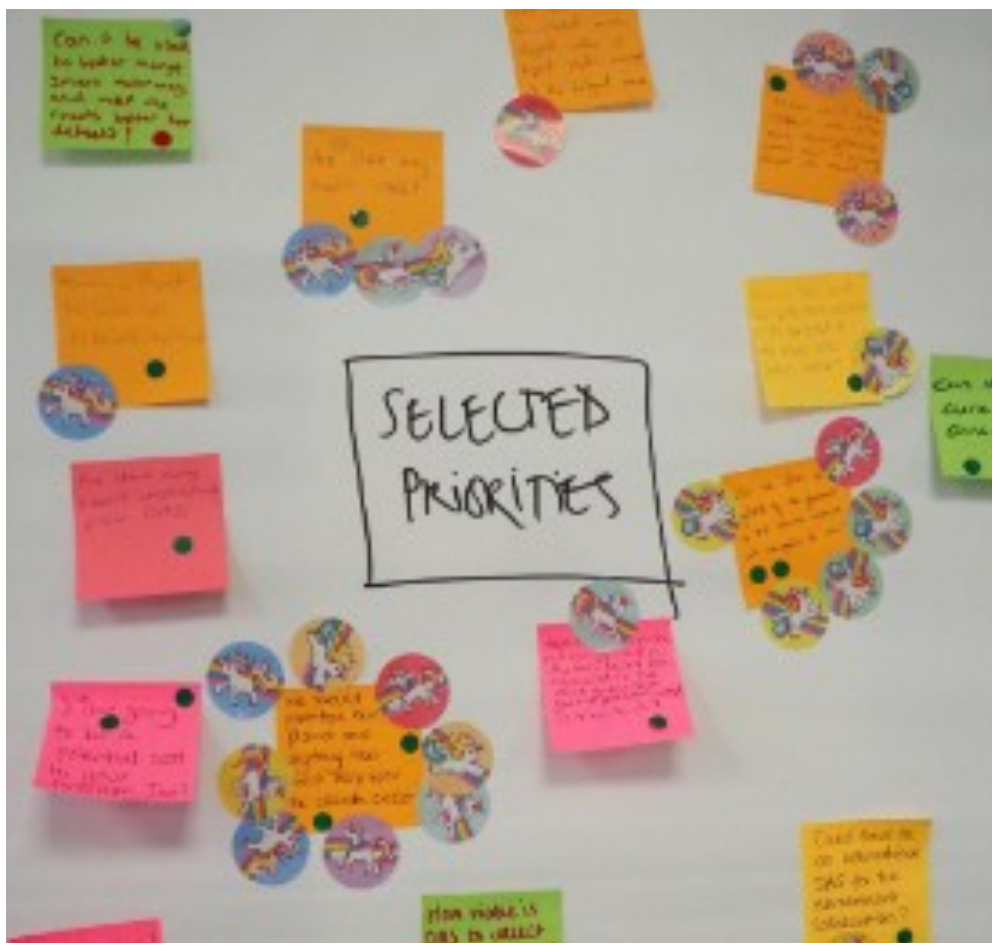
SELECTED PRIORITIES

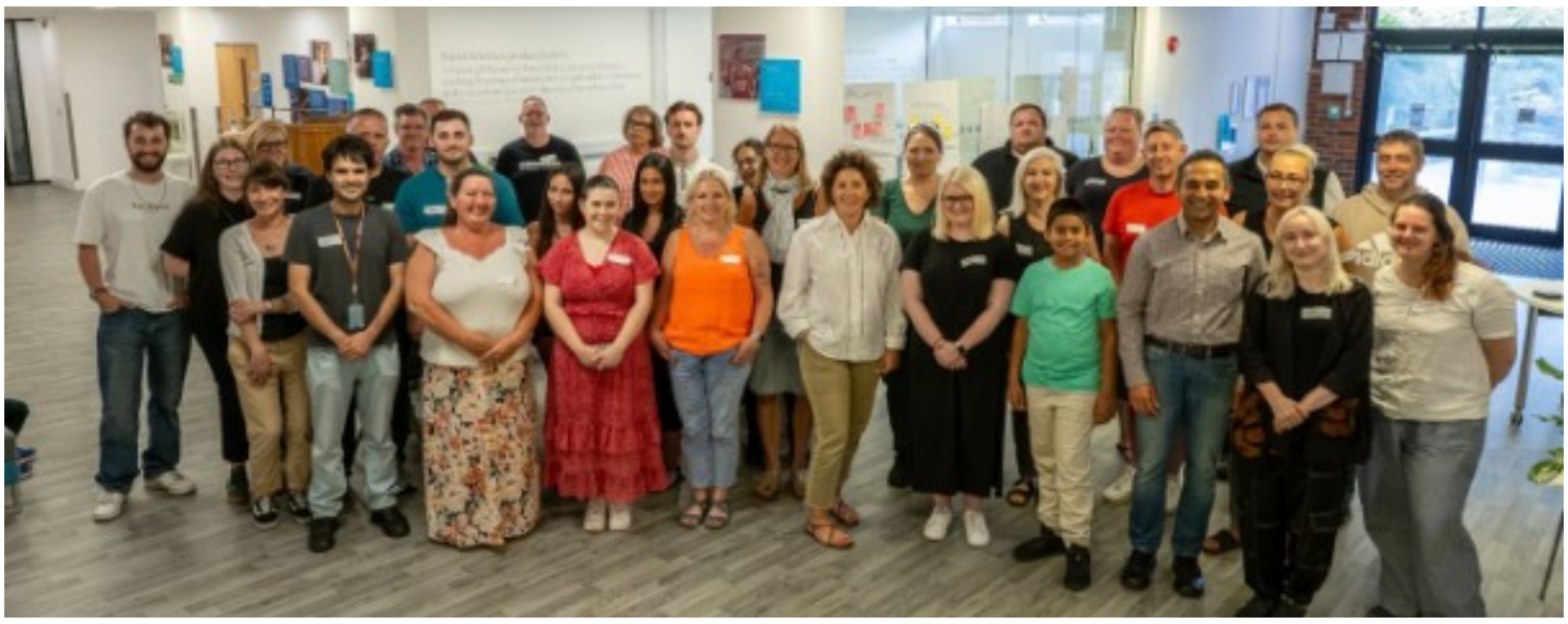

# Appendix

## Research ethics

The recruitment and design of the workshops for the Citizens' Panel were reviewed by NatCen's internal research ethics committee (REC). The REC consists of researchers who critically review all NatCen projects in line with the Government Social Research (GSR) Professional Guidance and the Economic and Social Research Council (ESRC) standards. The REC procedure provides ethical advice and guidance ahead of fieldwork, ensures that all research meets a high ethical standard, and provides reassurance to potential research participants. NatCen's REC process also reviews approaches to informed consent, the handling of participant data, and appropriate support offered to participants and researchers.

In reviewing this project, the REC committee agreed that the topic of DAS was not likely to cause members of the public undue distress and therefore it was not necessary to signpost additional support. The committee noted the substantial amount of time that participants were expected to engage with the topic; to mitigate this the research design involved regular breaks and active exercises in the final workshop including an outdoor activity. Given the sample focused on the general population and there were no quotas on children or vulnerable adults, the REC agreed that NatCen's standard safeguarding procedure should be followed with no requirement for additional safeguarding measures.

Additional ethics approval was obtained by the SOUNDSCALE team from the University of Southampton's Faculty of Engineering and Physical Sciences Research Ethics Committee (ERGO/FEPS/104481.A1). For those who consented, this allowed the SOUNDSCALE team to observe the panellists in a non-intrusive manner for research purposes, through video and audio recordings and direct observations.

## Panel recruitment

It was agreed that a panel of 40 citizens was the minimum size possible to bring together a diverse group of residents within the budgetary constraints for the project. To account for no-shows and ensure at least 40 participants completed all workshops, 50 participants were recruited. In total, 48 participants (23 in London and 25 in Hampshire) attended all three workshops. A full breakdown of the achieved sample is outlined in Table 1.

Participants were recruited through NatCen's trusted recruitment partner, Criteria Fieldwork, using a screening questionnaire. Participants were offered a total of £220 in cash for participating in all Panel sessions. The incentive was calculated on the basis of £60 per online session and £100 for the in-person workshop. Travel and caring expenses were reimbursed, and anyone experiencing barriers to online participation was supported through access to laptops, one-to-one Zoom support or reimbursing data charges. Criteria suggested that the incentive be paid in cash rather than other incentives like vouchers to maximize the chances of meeting quotas, especially for lower social grades.

Criteria used local recruiters to advertise the Citizens' Panel opportunity to residents via existing databases and relevant local forums. Members of the public interested in taking part contacted the local recruiter, who ran through a screening questionnaire to recruit against the quotas listed in Table 9. Anyone who had participated in any market or social research in the past 12 months was unable to take part. The recruiters also went through an information sheet to ensure informed consent.

# Panel sampling

Quotas were set to recruit a "mini-public" – a sample of the broader population – reflective of the demographic and attitudinal diversity of the two areas chosen (Hampshire and London). Quotas were defined according to a set of location, demographics, inclusion, attitudes, and pre-existing knowledge criteria, listed below. Postcode-cable proximity was the primary criterion and the remaining criteria were secondary, with flexibility given on a case-by-case basis during recruitment.

- **Location:** As a place-based Citizens' Panel, distance from residence to the cable location was the most important sampling criterion. Geographical boundaries were set around the optical fibre cable that the SOUNDSCALE team has access to through the National Dark Fibre Facility (NDFF). Using Geographic Information System (GIS) mapping software, a 500-metre boundary zone was created around postcodes in London boroughs (Figure 2), compared to a larger 2 km boundary zone in Hampshire, where the population density is much lower. Central London boroughs were excluded due to high concentrations of non-residential areas.
- **Demographics:** Using 2021 ONS census figures, demographic diversity was achieved through quotas on age, gender, social grade, ethnicity, education, disability and long-term illness. Quotas were increased for participants under 30 years old and those from racially minoritised groups, as previous dialogues and public opinion research demonstrate that these groups are more likely to drop out of deliberative processes.
- **Inclusion**: Quotas were increased for those from lower social grades, as their views are often underrepresented in social research. Those living close to existing 'dark fibre' networks in traditionally deprived areas and/or areas in cities with populations that experience marginalisation were identified as particularly important to ensure inclusion in this Panel. Thus, quotas were set using the Indices of Multiple Deprivation (IMD) from 2019 to classify relative deprivation (a measure of poverty) using UK Government data. This index considers several domains within its categorisation (income, employment, education, health, crime, barriers to housing and services, and living environment), making it a comprehensive index to use. Postcodes within the identified geographical boundary were assigned their associated deciles based on local super output areas (LSOA). Deciles range from 1 to 10, where decile 1 indicates the most deprived and decile 10 the least deprived. To increase representation from these underrepresented groups, a higher percentage of postcodes were sampled from deciles 1 and 2 (see Table 1). There were no households belonging to the 10$^{th}$ decile.
- **Attitudes**: A range of worldviews was captured using a spread of political leanings towards major parties in line with best practice (such as the recommendations of the UK Climate Assembly evaluation). Given DAS is an emerging technology with low public awareness, there is no trusted survey measure to capture a range of views towards DAS. Views towards privacy would not have been useful as a proxy as they only touch on one aspect of DAS and would not reflect a true spectrum of views.
- **Knowledge and experience of DAS:** To remove bias towards or against DAS technology and ensure views that reflected those of the general public, participants working in optical fibre technologies, but not those working in broader, related areas (for example, IT or telecommunications), were excluded. Participants knowledgeable about DAS or Distributed Fibre Optic Sensing (DFOS, a broader term which includes DAS and other sensing techniques) were also excluded.

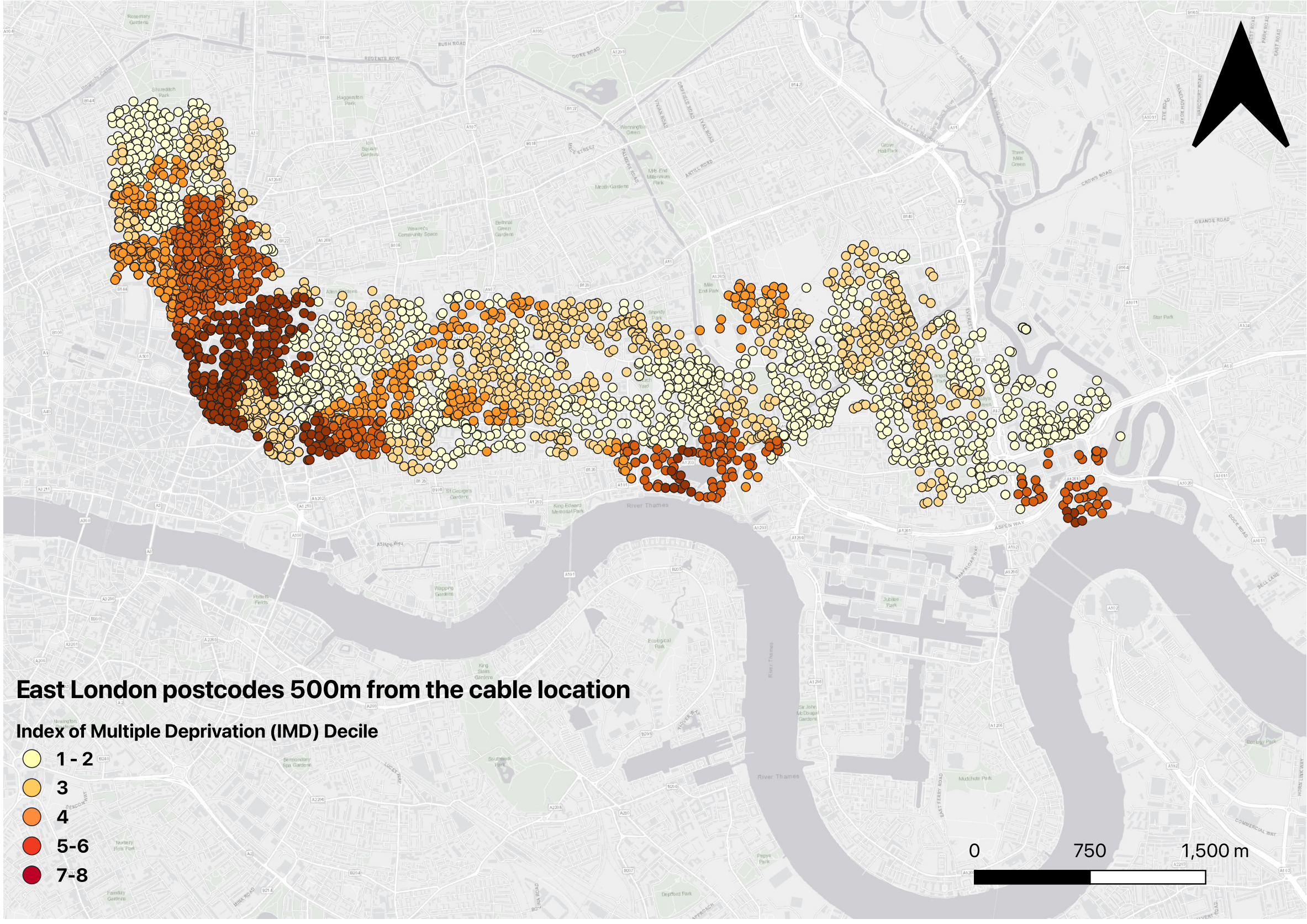

**Figure 3:** Example of maps produced to identify postcodes for citizen recruitment within a spatial buffer zone of 500 metres from the cable in London. Decile 1 indicates the most deprived and decile 8 the least deprived (there were no households from the 9th or 10th decile in this area). Map created in QGIS.

To support a comparison of views between London and Hampshire, we attempted to 'match' the sample so that the Panels were comparable demographically. The sample was stratified to reflect the local population of Southampton, and quotas for each sample criterion were created (see Table 1). These quotas were matched to the London boroughs by recruiting participants with similar demographics, rather than creating a separate, London-specific representative sample. This matched approach allows for more robust analysis of differences between the two locations. Any differences in views between similar types of people in the different locations could be better attributed to place. Due to small sample sizes in each location, the results are not statistically representative of the population, and so broader conclusions about the views of particular populations cannot be drawn.

Overall, the achieved sample reflected the Southampton population in key demographic characteristics such as gender, age, and disability, and included a spread of different political perspectives. There were marginally more women who attended than men, and there was an overrepresentation of 45–64-year-olds across both Hampshire and London. In Hampshire, desired quotas were not achieved for any ethnic minority groups, creating an overrepresentation of participants who identified as white. In both locations, quotas were not achieved for participants with no qualifications, and participants within AB and C1 social grades were overrepresented compared to C2 and DE. While the London sample achieved quotas for the Index of Multiple Deprivation, the Hampshire sample did not meet quotas for deciles 1 and 2 (highest deprivation).

A total of 48 participants completed all three workshops, which is 20% greater than the target sample size of 40. This over-recruitment is likely due to the strong networks of recruiters in London and Southampton, who drew on local relationships and advertised in relevant local forums to recruit. The additional participants tended to be from higher social grades, however, since some of the lower social grade quotas were not met. This was partly due to a smaller proportion of lower-decile postcodes available close to the cable in Hampshire. It also reflects wider challenges in engaging people from lower social grades in research, which may have been heightened by the technical nature of the topic. Although a high cash incentive was offered to overcome some of these barriers, there may be lessons to learn about how the project is framed and communicated to better engage underrepresented members of society. Good representation across educational attainment quotas was achieved, however.

# Quotas

**Table 9:** Sampling criteria and targets (Hampshire and London).

| Criteria | Categories | % of the Southampton population[21] | Target (total 20) | Achieved (Hampshire, total 25) | Achieved (London, total 23) |
|---|---|---|---|---|---|
| **Gender** | Man | 48.5-49%[22] | 9 | 11 | 10 |
| | Woman | 51% | 10 | 14 | 13 |
| | Non-binary / other / prefer not to say | 0.5% | 1 | 0 | 0 |
| **Age**<br><br>*Overrecruited for participants under 30* | 18-29 | 21% | 7 | 6 | 7 |
| | 30-44 | 22% | 5 | 5 | 6 |
| | 45-64 | 18% | 4 | 13 | 7 |
| | 65+ | 14% | 4 | 1 | 3 |
| **Ethnicity**<br><br>*Overrecruited for racially minoritised participants* | White | 80% | 13 | 23 | 11 |
| | Asian or Asian British | 11% | 3 | 1 | 3 |
| | Black, Black British, Black Caribbean or Black African | 3% | 2 | 1 | 5 |

[21] 'Southampton' here refers to the Southampton local authority district.

[22] Exact percentage to be recruited conditional on success of recruitment of a participant who identifies as non-binary, or who states 'other' or 'prefer not to say' with regard to gender identity.

| | | | | | |
|---|---|---|---|---|---|
| | Mixed or multiple ethnic groups | 3% | 2 | 0 | 3 |
| | Other ethnic group | 2% | 1 | 0 | 1 |
| **Disability / LTI** | Yes | 20% | 4 | 3 | 4 |
| | No | 80% | 16 | 22 | 19 |
| **Education** | Level 4 qualifications or above[23] | 31% | 6 | 8 | 7 |
| | Level 3 qualifications[24] | 21% | 4 | 3 | 5 |
| | Apprenticeship | 5% | 1 | 0 | 1 |
| | Level 2 qualifications[25] | 13% | 3 | 7 | 5 |
| | Level 1 qualifications[26] | 10% | 2 | 6 | 1 |
| | Other[27] | 3% | 1 | 0 | 2 |

[23] Level 4: Degree (BA, BSc), higher degree (MA, PhD, PGCE), NVQ level 4 to 5, HNC, HND, RSA Higher Diploma, BTEC Higher level, professional qualifications (for example, teaching, nursing, accountancy).

[24] Level 3: 2 or more A levels or VCEs, 4 or more AS levels, Higher School Certificate, Progression or Advanced Diploma, Welsh Baccalaureate Advance Diploma, NVQ level 3; Advanced GNVQ, City and Guilds Advanced Craft, ONC, OND, BTEC National, RSA Advanced Diploma.

[25] Level 2: 5 or more GCSEs (A* to C or 9 to 4), O levels (passes), CSEs (grade 1), School Certification, 1 A level, 2 to 3 AS levels, VCEs, Intermediate or Higher Diploma, Welsh Baccalaureate Intermediate Diploma, NVQ level 2, Intermediate GNVQ, City and Guilds Craft, BTEC First or General Diploma, RSA Diploma.

[26] Level 1: 1 to 4 GCSEs grade A* to C, Any GCSEs at other grades, O levels or CSEs (any grades), 1 AS level, NVQ level 1, Foundation GNVQ, Basic or Essential Skills.

[27] Other: vocational or work-related qualifications, other qualifications achieved in England or Wales, qualifications achieved outside England or Wales (equivalent not stated or unknown).

| | | | | | |
|---|---|---|---|---|---|
| | No qualification | 17% | 4 | 1 | 2 |
| **Social grade** | AB (Higher managerial, administrative, and professional occupations) | 18% | 4 | 8 | 7 |
| | C1 Supervisory, clerical and junior managerial / administrative / professional | 33% | 6 | 9 | 8 |
| | C2 Skilled manual occupations | 23% | 5 | 4 | 3 |
| | DE Semi-skilled and unskilled manual occupations; unemployed and lowest grade occupations | 26% | 5 | 4 | 5 |
| **Party identification** *Q: Do you think of yourself as a little closer to one political party than to the others? If yes, which one?* | Conservative | N/A | 3 | 1 | 4 |
| | Labour | N/A | 3 | 1 | 8 |
| | Liberal Democrat | N/A | 3 | 6 | 3 |
| | Green Party | N/A | 3 | 2 | 2 |
| | Reform | N/A | 3 | 3 | 2 |
| | None | N/A | 5 | 12 | 4 |
| **Knowledge and experience of DAS** | Familiar with DAS/DFOS | N/A | TBD on a case-by-case basis | 0 | 0 |
| | Works in the field of optical fibre technologies | N/A | 0 | 0 | 0 |
| **IMD via postcode proxies – London and Hampshire** | Decile 1 | 20% | 4 from corresponding postcodes | 0 | 3 |
| | Decile 2 | 20% | 4 from corresponding postcodes | 1 | 7 |
| | Decile 3 and 4 | 20% | 4 from corresponding postcodes | 3 | 4 |
| | Decile 5-10 | 40% | 8 spread evenly across corresponding postcodes | 21 | 9 |

# Research priority voting

At the beginning of the final workshop, participants were asked to come up with specific questions regarding the five thematic areas that they would be interested in the SOUNDSCALE team to research on. Here's a list of all the questions participants came up with, divided into which area they were categorised in.

**Table 10:** Final list of questions put forward for voting at the end of the Panel.

| Area and questions (London) | Area and questions (Southampton) |
|---|---|
| **City Infrastructure** | |
| How will DAS get on with other cable providers? (Cable Sharing) | Can the DAS fibres be re-purposed to aid better broadband etc? |
| How does DAS interact with other technologies? | What is the sensitivity of the equipment? |
| The cable east to west is a test run or is it the main hub? | Underground leaks? |
| | Can this be used to identify if an area is overpopulated? People and buildings? |
| **Surveillance and Crime** | |
| Could it be used for voice recognition and therefore used as evidence for voice recognition? | I would like to know more about tracking of people and conversations |
| What are the ethics of DAS? | How many cables and therefore detectors would be needed per city for monitoring? |
| Will it be listening always? | Safety alerts for young people |
| Who has control? | If you can deduce who is walking down the street, it would be easy to disguise that gait by changing walking style or riding a scooter to get around. |
| How would DAS affect on surveillance/privacy of civilians? | Why try and detect fly tipping if visual evidence is needed to take them to court? |
| How precise can it be? Will evidence meet a criminal standard? | |
| Can it support crime data collection and support better fund allocation? | |
| Crime equality, if only deployed in high crime areas? | |
| Can police track DAS data? | |
| How do you know who owns parts of DAS tech? Outputs (data)? Infrastructure? | |
| **Governance and Costs** | |
| Who are the stakeholders? (from top to bottom) | What regulations are already in place for controlling DAS? |
| Ownership to implement it. Central or local government. Private or Public. | Is there a guarantee that it will be prioritised for good uses rather than 'corporate'? |
| What is / is there a cost to us? | Is this a complex solution when other tech can do the job better? |
| Where is the money coming from (Private or Public)? | How does DAS technology compare cost wise to other existing technologies? |
| What are citizens rights of this being implemented in their area? Is there a mandatory citizen consultation? | Who would be paying for the cost of it? e.g. health - tax? |
| How will the contractor be procured? Open tender? Existing framework? | How can the data be analysed when it is so big? Especially by companies that may buy access. |
| How can we give people more power to have a choice over this? | Lets not introduce more risk into our world. SAFE technology only - don't invent more invasiveness. |
| Is there costs involved to the public? | Who is inputting this data? What experience do they have? Knowledge wise? |
| Should DAS be public or private? (it's deployed in public, so should public own it?) | Different types of fibreoptic cables - would they need to lay new cables / change them? |
| How will DAS be funded? By who? | How can we ensure DAS will be used for good when the cables are owned by unregulated, private companies? |
| Short term 'non-extreme' cases in the UK. What are they thinking? | How can we persuade the government? |
| How do we protect/manage/opt out to our data being used? | How will this be funded and will the costs be passed on to us? |
| Who can 'buy' into it? | How would the government be able to access private company's existing use of DAS? |
| How is it going to be monitored long term? Will the right stakeholders be involved? | How can we get the government to listen to concerns regarding DAS? |
| Vest interests? Special interest groups? | Just how much might it cost to monitor foot steps around a city and therefore track criminals? |
| Risks of scam with the data, how will it be controlled? | |
| If we can upgrade DAS for the future? | |
| How can we hold parties / stakeholders accountable? (Licensing system?) | |
| Is it central or local government funded? And which department is responsible for it? | |
| Will this research properly get into things such as value for money or is it more focussed on what is scientifically possible? | |
| Is the funding of research justified? | |
| Who currently uses DAS and for what purpose? | |
| Can we get brands or sponsors to pay for the research? | |
| How to decide when DAS research benefits outweigh the costs? | |
| Is it private or public owned? | |
| Who will take responsibility / Who are they accountable to? | |
| Who would be held accountable if DAS went wrong? | |
| Will laws change to accommodate or will existing laws protect? | |
| How do you monitor that the project is still viable? | |
| What does success look like and how do you know it its working? | |
| Lightly regulated and checked – utilising both human creativity and ingenuity and robotic logic | |
| Do the 'dark' cables not belong to an owner? Who is likely to want the cable returned to original uses as data volumes increase in the future? | |
| **Public Health and Community** | |
| Why has this come about now? How can this really support and help health? (most important area) | I was thinking about the health and safety |
| What the process to deploy DAS in cities / local areas? | Can DAS improve quality of life by reducing pollution + diverting traffic? |
| How can we ensure this technology will not discriminate against disadvantage groups? | Can it help emergency services respond quicker? |
| Will the result of using DAS make it easy for the public to understand how this will help them? | Can DAS make appointments come quicker with data? |
| How do we help people understand DAS? (Digital literacy) | |
| Who is going to deal with 'customers'? | |
| How to incorporate DAS health research into wider research frameworks? | |
| Regulate the comms. | |
| Regulate claims of speeds of broadband. | |
| Policing high levels. What service for DAS to give up for comms. | |
| **Environment** | |
| DAS – how will it improve environmental pollution? | Detecting an overall noise level due to insect life may be possible per sq. meter but would it be possible over a larger area? |
| What's the best way to maintain DAS? | Is there a known upper limit at which DAS proves ineffective? For example, would it be possible to monitor an ant farm next to a music festival? |
| How does background vibrations access the sensitivity and use of the data collected? | How would smart cities benefit health and the environment? |
| How would it affect the environment? | How can DAS be used to prevent natural disasters? Cost? Reliability? Risks? |
| | More scientific information what/how is it currently used? |
| | How will this affect us? |
| | If the cables are already in place, how badly does it affect the environment? |
| | How can we ensure minimal environmental damage or prevent it when it comes to the use of DAS? |
| | Can this be used to identify different species of animals? |
| | Agriculture use e.g. monitor soil health? |
| | Would the benefits/uses outweigh the environmental costs? i.e. Data centre usages? |
| | Can DAS be used to detect pollution? Weight / Fluid levels / Vibrations etc |
| **Something Else** | |
| Regarding the different areas have the specialists in these areas been consulted? | How exactly does DAS use the fibre optic cables to detect vibrations, such as traffic on roads or tectonic shifts in the ocean? |
| How do we know there will not be another emerging technology which can help to prevent the need to dig to pit additional cables for proposed use of DAS? | DAS is already out there, why reinvent the wheel? |
| Is this technology completely new or is it properly implemented in any countries? | Interference of cables e.g. AC, TV, etc. |
| What is DAS's current capability? | We have a lot of entertainment using internet, streaming etc. which is very quick, can this tech |
| | make it even easier./quicker to download films, music, stream etc? |
| Were we guided towards these chosen areas? | Will large scale events (Boomtown etc.) overload the ability to get useful readings? |
| Is this technology needed in built up cities? | Marginalise poor sections of society? |
| How to anonymise DAS data. | Are we going to become reliant on it? |
| If it is for detecting earthquakes in countries prone to earthquakes. | Is DAS being used in any way in the UK currently? |
| Multiple-Flat buildings vs individual homes? | Will the info gleaned from the Citizens' Panel inform or dictate future projects? |
| DAS smart readings? | How sensitive and how decipherable is the audio & measurement of it? |
| How sensitive? Can a particular person be tracked? Identified gait analysis? Can a cat be tracked? And at what cost? | Will the owners of the cables engage correctly? or will they dictate use / be hard to work with? |
| How long till you decide DAS isn't a viable technology? | Are you in contact with WinAcc? (Winchester against climate change) |
| Where is DAS being tested in cities? Or where will it be tested? | Time specific capabilities on data retrieval. |
| Can DAS detect faults in planes (before take off), bridges, roads? | Storage of data and environment possibilities from the data centre such as energy re-use. |
| | Who will make the decision on what data will be collected? |
| | How will we know if DAS is being used? And who by? |

At the end of the final day, a few of the previous question (including new ones that they added throughout the day as they continued with their deliberations, discussions and activities), participants were given the chance to vote on which individual priority they would like the SOUNDSCALE team to take forward to investigate. The following two tables summarise the questions chosen by each location, and the number of 'stars' each priority received.

**Table 11:** Final voting for priorities by the panellists in Southampton.

| Area and questions (Southampton) | Starred |
|---|---|
| **Environment** | |
| Having a planet to live in is more important. | 1 |
| We should prioritise our planet and anything that could help with the climate crisis | 8 |
| Could there be an international DAS for the environment collaboration? | 1 |
| **Public health and community** | |
| Are there any health uses? | 3 |
| Can DAS cure for cancer? | |
| Are there any health implications from DAS? | |
| How viable is DAS to collect population level data considering the data and analytical challenges? | |
| **Surveillance and crime** | |
| Can we track an individual across different cables at depths amongst all the background noise? | 1 |
| How will this affect us with the surveillance and our privacy that we need? | 3 |
| **City infrastructure** | |
| Can it be used to better manage smart motorway and map the roads better for defects? | |
| **Governance and costs** | |
| How can DAS be regulated by the government so that private companies can't monopolise its use? | 5 |
| Is there going to be a potential cost to your average Joe? | |
| Have the cost implications and energy usage for the data centres been calculated vs the value of the data gained/potential usages? Is it worth it? | 1 |
| How will irrelevant data be destroyed? | |
| How will the government be able to regulate DAS effectively? | |
| What can you do to make the government to understand the important of DAS? | |
| **Something else** | |
| How is DAS (and our potential reliance on it) affected if the fibre optic cables break? | 1 |
| How fast can the data be processed? | 2 |

**Table 12:** Final voting for priorities by the panellists in London.

| Area and questions (London) | Starred |
|---|---|
| **Environment** | |
| What impact does it have on wildlife habitats | |
| What is the environmental impact on the placement of the data centre to the surrounding areas? | |
| Does DAS track why is there is a lot of pollution near airports? | |
| **Public health and community** | |
| Can DAS be used to produce a noise scale in areas where people live? To judge how noisy an area is to live in? | |
| **Surveillance and crime** | |
| [None] | |
| **City infrastructure** | |
| How will this impact the people in my areas being placed? | |
| **Governance and costs** | |
| How can DAS be regulated and by who? | |
| Who owns the data? | |
| If the project is not longer viable will it be stopped/closed? At what point will viability be checked? | |
| Would the data prevent or report on findings? Would it be beneficial in the long run? | |
| How can we regulate the people/users using the technology – Licensing? | 2 |
| Who is currently interested, investing and researching the DAS technology? | 1 |
| How do we monitor and regulate through its evolution? | 5 |
| Is there a national security risk? | |
| Cost vs benefit of the project | 5 |
| DAS = regulating. Facebook/TikTok – used for the wrong purposes | |
| Who has a 'licence to operate'? | |
| Can footfall data be sold to companies looking to choose a retail location | |
| **Something else** | |
| What are the disadvantages of not using DAS? | 5 |
| Why? | 2 |
| If they can add on tech rather than just pick up sound? | 1 |
| What are the advantages of going first? | |
| Can the cable be used for multiple areas to address? Regarding different locations and every multiple research area? | 1 |
| Direct tax or business payment? | |